\documentclass[aps,pra,reprint,superscriptaddress]{revtex4-2}

\usepackage[T1]{fontenc}
\usepackage{lmodern}
\usepackage[french,english]{babel}
\makeatletter
\@ifundefined{l@fre}{\let\l@fre\l@french}{}
\@ifundefined{captionsfre}{}{}
\@ifundefined{datefre}{}{}
\@ifundefined{extrasfre}{}{}
\@ifundefined{noextrasfre}{}{}
\makeatother
\usepackage{amsmath,amssymb,amsthm,mathtools}
\usepackage{bm}
\usepackage{graphicx}
\usepackage{xcolor}
\usepackage{physics}

\usepackage[colorlinks=true,linkcolor=teal,citecolor=teal,urlcolor=teal,plainpages=false,pdfpagelabels]{hyperref}

\usepackage{tikz}
\usepackage{times}

\newcommand{\nocontentsline}[3]{}
\let\origcontentsline\addcontentsline
\newcommand\stoptoc{\let\addcontentsline\nocontentsline}
\newcommand\resumetoc{\let\addcontentsline\origcontentsline}

\newcommand{\bbC}{\mathbb C}

\newcommand{\cA}{{\mathcal A}}
\newcommand{\cB}{{\mathcal B}}
\newcommand{\cC}{{\mathcal C}}
\newcommand{\cD}{{\mathcal D}}

\newcommand{\cF}{{\mathcal F}}
\newcommand{\cG}{{\mathcal G}}
\newcommand{\cH}{{\mathcal H}}

\newcommand{\cK}{{\mathcal K}}
\newcommand{\cL}{{\mathcal L}}
\newcommand{\cM}{{\mathcal M}}
\newcommand{\cN}{{\mathcal N}}
\newcommand{\cO}{{\mathcal O}}
\newcommand{\cP}{{\mathcal P}}

\newcommand{\cR}{{\mathcal R}}
\newcommand{\cS}{{\mathcal S}}

\newcommand{\cU}{{\mathcal U}}
\newcommand{\cV}{{\mathcal V}}

\newcommand{\cZ}{{\mathcal Z}}

\renewcommand\a{\alpha}

\renewcommand\c{\chi}

\renewcommand\d{\delta}
\newcommand\D{\Delta}
\newcommand\e{\epsilon}
\newcommand\ve{\varepsilon}

\newcommand\G{\Gamma}
\newcommand\h{\eta}

\renewcommand\l{\lambda}
\renewcommand\L{\Lambda}

\newcommand\m{\mu}
\newcommand\n{\nu}
\newcommand\p{\phi}
\renewcommand\P{\Phi}

\renewcommand\r{\rho}
\newcommand\s{\sigma}
\renewcommand\S{\Sigma}
\renewcommand\t{\tau}

\newcommand\w{\omega}
\newcommand\W{\Omega}
\newcommand\x{\xi}
\newcommand\y{\psi}

\newcommand{\id}{\mathbf 1}

\providecommand{\ket}[1]{\lvert #1\rangle}
\providecommand{\bra}[1]{\langle #1\rvert}
\providecommand{\Tr}{\operatorname{Tr}}

\newcommand{\Stab}{\operatorname{Stab}}
\providecommand{\abs}[1]{\lvert #1\rvert}
\providecommand{\norm}[1]{\lVert #1\rVert}
\newif\ifhascitations
\hascitationstrue
\newcommand{\thmref}[1]{\textnormal{[#1]}\ }

\newtheorem{theorem}{Theorem}
\newtheorem{proposition}[theorem]{Proposition}
\newtheorem{lemma}[theorem]{Lemma}
\newtheorem{corollary}[theorem]{Corollary}

\theoremstyle{definition}
\newtheorem{definition}[theorem]{Definition}
\theoremstyle{remark}

\begin{document}
\stoptoc

\title{Infinite magic and the emergence of the continuum}

\author{Leo Shaposhnik}
\thanks{First and corresponding author.
Email: \href{mailto:leoshaposhnik@gmail.com}{leoshaposhnik@gmail.com}.}
\affiliation{Dahlem Center for Complex Quantum Systems, Freie Universit\"at Berlin,  
14195 Berlin, Germany}

\author{Lorenzo Leone}
\affiliation{Dipartimento di Ingegneria Industriale, Universit\`a degli Studi di Salerno, %Via Giovanni Paolo II 132, 
84084 Fisciano (SA), Italy}
\affiliation{Istituto Nazionale di Fisica Nucleare (INFN) Sezione di Napoli, Gruppo Collegato di Salerno, Italy}
\affiliation{Dahlem Center for Complex Quantum Systems, Freie Universit\"at Berlin, 
14195 Berlin, Germany}

\author{Lennart Bittel}
%\affiliation{Department of Physics, Institute for Theoretical Physics, Freie Universit\"at Berlin, 
%14195 Berlin, Germany}
\affiliation{Dahlem Center for Complex Quantum Systems, Freie Universit\"at Berlin,  
14195 Berlin, Germany}

\author{Dimitris Saraidaris}
%\affiliation{Department of Physics, Institute for Theoretical Physics, Freie Universit\"at Berlin, 
%14195 Berlin, Germany}
\affiliation{Dahlem Center for Complex Quantum Systems, Freie Universit\"at Berlin, 
14195 Berlin, Germany}

\author{Alexander Jahn}
\affiliation{Dahlem Center for Complex Quantum Systems, Freie Universit\"at Berlin, 
14195 Berlin, Germany}
\affiliation{Asia Pacific Center for Theoretical Physics, 37673 Pohang, Gyeongsangbuk-do, South Korea}

\author{Jens Eisert}

\affiliation{Dahlem Center for Complex Quantum Systems, Freie Universit\"at Berlin, 
14195 Berlin, Germany}

\date{\today}

\begin{abstract}
Which features of a quantum state are needed for a relativistic quantum field theory to emerge from a lattice system? We address this question by establishing 
a quantitative relationship between the type classification of hyperfinite von Neumann algebras and the quantum-information-theoretic resource
of nonstabilizerness, also called magic. Our results establish two complementary
parts of this relationship. First, we prove that any stabilizer system on a lattice defines a canonical GNS Hilbert space in which local von Neumann algebras do not contain type III factors in their central decomposition.
%\LS{the trace over the codespace defines a state on the infinite-dimensional algebra of quasi-local operator and thus leads to a well-defined GNS Hilbert space associated to the infinite-dimensional codespace. We then establish that in this Hilbert space, the von Neumann algebra associated to any subregion of the lattice contains  no type III factor in its central decomposition. 
As an explicit example, we prove that for stabilizer systems with short-range order, the
local algebras are isomorphic to tensor products of hyperfinite factors of type I
or type II. This demonstrates that the absence of magic is accompanied by the absence of type III factors in the local algebras, therefore prohibiting the appearance of a quantum field theory. Second, for the hyperfinite factors of Araki, Woods and Powers, %\je{bipartite infinite tensor products of finite type-I factors}
we prove a quantitative equivalence: The limiting factor is of type
III if and only if the locally optimized min-relative entropy of magic
diverges. This divergence is necessary to generate the continuous spectrum associated to a type III factor.
Thus, in this setting, the emergence of a type III factor requires not merely
the presence of magic, but an infinite amount of it. % of magic that grows without bound. 
These results provide a rigorous setting supporting the proposal that infinite magic
is a necessary structural ingredient for the emergence of a relativistic
continuum from lattice systems.
\end{abstract}

\maketitle
\section{Introduction}

The discovery of black-hole thermodynamics \cite{Bardeen:1973gs,Hawking1975}, the generalized
second law \cite{Bekenstein:1974ax} and the area law of
entanglement entropy \cite{Bombelli:1986rw,Srednicki:1993im,Eisert2010} 
have impressively 
demonstrated that notions of quantum information
theory and the study of spacetime geometry are intimately related. The AdS/CFT
correspondence \cite{tHooft:1993dmi,Susskind:1994vu,Maldacena:1997re,Witten:1998qj} has given this
relationship a particularly %sharp 
crisp form: One modern reading of this correspondence is that the bulk spacetime geometry 
emerges from the quantum-information-theoretic structure of the dual boundary
state \cite{Ryu:2006bv,Hubeny:2007xt,VanRaamsdonk:2010pw,Lewkowycz:2013nqa,Faulkner:2013ana,Jafferis:2015del}. In this context, the relationship between entanglement entropy and geometry is
captured
by the Ryu-Takayanagi-formula \cite{Ryu:2006bv}, which relates the total entanglement entropy of a
subregion $A$ to the area of a surface that bounds the entanglement wedge, the region of the bulk geometry which emerges from $A$. These observations led to a collection of studies which model the process of bulk-boundary reconstruction in AdS/CFT using the language of quantum error correction~\cite{Almheiri:2014lwa, Dong:2016eik, Harlow:2016vwg, Pastawski:2015qua, Jahn2019, Cree2021, AnglesMunne2022, Steinberg2024}.
This intertwinement of the study of spacetime geometry and quantum fields with the mindset of quantum information science, 
driven by notions of information processing, provides a compelling research perspective.

More recently, a novel perspective on this relationship has been
emerging that emphasizes the \emph{structure of entanglement} rather than its absolute amount~\cite{cao2024gravitationalbackreactionmagical,Cao:2026uoq,Tirrito:2023fnw,Benedetti:2026mfy,Moosa:2026nqz}. One of the 
insights of this perspective is that the proper incorporation of gravitational dynamics requires the information-theoretic measure of entanglement known as \emph{nonstabilizerness} or \emph{magic}
\cite{Emerson:2013zse,Bravyi:2018ugg}.
This quantity is important 
as a resource that conceptually quantifies how far a quantum state departs from a stabilizer description; in quantum computing, it takes center stage in notions of quantum error correction \cite{Roads}.
Common toy models \cite{Pastawski:2015qua, Jahn2019, Cree2021, AnglesMunne2022, Steinberg2024}
of the correspondence are based on \emph{stabilizer quantum error correction,} in which magic is entirely absent. This raises the question \cite{Pastawski:2015qua}:
\begin{quote}
    Can a relativistic quantum field theory emerge from a stabilizer system or is this impossible?
\end{quote}
In this work, we turn this question 
into a quantitative connection between the operator-algebraic structure associated
with a relativistic quantum field theory
\cite{Fredenhagen1985} and the presence of magic in the underlying system. 
Our results separate two questions that
are closely related but logically distinct. We first ask which von Neumann algebras can arise from
an exact stabilizer system on an infinite lattice and whether this resembles the type III$_1$ algebras appearing in relativistic quantum field theories. After establishing the incompatibility of these two notions, we then ask how the amount
of magic must scale along a sequence of finite systems in order for a type III
factor to emerge in that limit. %The first question leads to a general sector-wise exclusion of type III factors; 
%the second leads, for a class of infinite tensor-product factors, 
For the hyperfinite factor of Araki-Woods \cite{ArakiWoods1968} and Powers \cite{Powers1967} this leads to an exact equivalence between
type III structure and divergent magic. We begin by sketching the intuition underlying these results, followed by a detailed formulation of our results in Sec.~\ref{sec:results} and their discussion in Sec.~\ref{sec:discussion}.

\subsection{A link between magic and quantum field theory}
For a quantum mechanical system consisting of qubits,
\begin{equation}
  \cH=\bigotimes_{i=1}^{N}\mathbb{C}^{2},
\end{equation}
with the set of Pauli operators denoted by $\cP$, a consistent stabilizer system $\cS \subset \cP$ is a set of commuting Pauli operators whose
finite products do not equal $-\id$.  Since they 
commute, these stabilizers define a subspace $C$ of the full Hilbert space as the
common $+1$ eigenspace of the stabilizer operators. This underlies the formalism of stabilizer quantum error correction \cite{Gottesman:1997zz}.  There the subspace $C$ is usually called the \emph{code subspace} or \emph{logical space}, whereas the full Hilbert space $\cH$ is considered the \emph{physical space}. This terminology arises from the fact that $\cH$ usually describes the full Hilbert space that constitutes a quantum computer and $C$ describes a Hilbert space of logical qubits encoded in it.

If on a Hilbert space with $N$ qubits one has $N$ stabilizers, the code space is uniquely fixed to contain only a single vector, a so-called stabilizer state. 
For this vector, the reduced density matrix for any subsystem
is fixed to the form 
\begin{equation}\label{eq:projection_state}
  \rho_{\Gamma}=\frac{P_{\cS}}{\operatorname{rank}P_{\cS}},
\end{equation}
where $P_{\cS}$ is a projection \cite{Fattal:2004frh}.
The density matrix of such stabilizer states therefore has a flat spectrum.
More generally, all vectors that arise from acting with \emph{Clifford circuits} (circuits generated by Pauli operators, Hadamard, CNOT, and phase gates) on a product vector in the computational basis are stabilizer states. The Gottesman-Knill theorem~\cite{gottesman1998heisenberg} asserts that unitaries that arise from  Clifford circuits can be efficiently classically simulated. Quantum circuits which cannot be simulated classically therefore require a significant chunk of non-Clifford unitaries.

%\lb{this seems \je{awkwardly} formulated. Maybe more: "\je{Operations} with Cliffords can be \je{simulated with} GK". "Quantum gates which cannot be simulated classically therefore require a significant chunk of \je{non-Clifford} operations...". \je{So} \je{Cliffords} are just a class, and there are others that can also be simulated} \LS{Now better?}

This insight has profound implications for our study at hand: Consequently, a pure state with a non-flat reduced spectrum cannot be a stabilizer state and cannot be prepared from a computational-basis vector using only Clifford gates. Its nonstabilizerness is called \emph{magic} and can be quantified by suitable \emph{magic monotones}. In particular, a non-flat reduced spectrum witnesses magic, although a flat reduced spectrum does not imply its absence.

Within quantum information theory, magic is
on the one hand seen
as a resource that allows one to estimate how hard a unitary is to implement on a classical computer and thus is not obviously related to the properties of systems found in nature
\cite{PhysRevLett.116.250501}. 
On the other hand, it plays an important role in quantum error correction \cite{Roads}: Clifford circuits can in many situations be transversally implemented and are hence relatively ``cheap'', while the uplifting to universal quantum circuits requires the addition of magic---inserted into the system by complicated magic state distillation \cite{PhysRevA.71.022316,PhysRevA.95.022316} or cultivation schemes \cite{Cultivation}. While this is an important connection and magic plays an important role within the 
stabilizer formalism of quantum error correction \cite{Gottesman:1997zz}, viewed from this perspective,
%In particular the standard framework to formulate quantum-error correction, the stabilizer formalism \cite{Gottesman:1997zz}, is mostly concerned with the problem of representing information within a quantum computer and}
it is not obvious whether magic is relevant in systems that arise naturally. 

Recent work, however, has shifted the perspective here: It has demonstrated the ability of certain magic measures to identify the critical point of many-body systems, which therefore suggests a deeper relevance of this concept for our understanding of nature. For the transverse-field Ising model, it has been
demonstrated that a specific magic measure, the \emph{stabilizer R\'enyi entropy},
peaks exactly at the critical
point~\cite{oliviero2022magicstateresourcetheory}. As is well known \cite{Pfeuty:1970qrn,LiebSchultzMattis1961,BloteCardyNightingale1986,Affleck1986}, the system can at its critical point be described in the infrared by a conformal
field theory (CFT), the $c = \frac{1}{2}$ minimal model. Similarly, a universal dependence of such measures on the subsystem size in a CFT has been derived using path integral methods \cite{Hoshino:2025jko}.
Furthermore, it has been argued that the existence of non-flatness of the
underlying state can, in a system that has a holographic dual, be linked to the
tendency of geometry to back-react onto matter~\cite{cao2024gravitationalbackreactionmagical}.
These observations point towards some role played by magic in the emergence of
a relativistic continuum, but this relationship is so far poorly
understood.
In Ref.~\cite{Chemissany:2025vye} it has been suggested 
that due to their flat spectrum, stabilizer states cannot lead to the emergence of so-called type III von Neumann algebras in infinitely large systems. These algebras universally appear as the local algebras in quantum field theory~\cite{Fredenhagen1985}. It is natural to expect that any finite injection of magic through non-Clifford resources in the preparation circuit should not change the large-scale entanglement structure of the state and still not lead to type III algebras. A natural proposal is then:
\begin{quote}
    Any state on an infinite lattice system that allows for an effective description in terms of a relativistic quantum field theory has infinite magic.
\end{quote}
Based on this it seems that infinite magic is, just as infinite entanglement, a necessary condition for a relativistic
continuum to emerge at large scales. The above proposal is stated heuristically and does not provide a formal theorem that one can prove, as the notion of effective description is not defined, nor are the systems constrained. However, an answer to this question would shed direct light on the question raised in the previous section. The purpose of this work is to isolate precise settings in which substantial parts of this proposal can be established rigorously. Our first main result concerns exact stabilizer systems on countable lattices of qubits. We prove, sector by sector, that no type III factor can occur in the central decomposition of their local algebras. Under an additional short-range-order condition, we obtain the stronger structural conclusion that the local algebras decompose into tensor products of hyperfinite factors of type I or type II.

Our second main result is quantitative. We study the hyperfinite factors introduced by Araki, Woods \cite{ArakiWoods1968}, and Powers \cite{Powers1967}, which are based on sequences of vectors that describe an infinite tensor product of entangled qubit pairs. We prove that the limiting factor is of type III if and only if the locally optimized min-relative entropy of magic in the underlying state diverges. Thus, within this family, it is not enough for magic merely to be present; its amount must grow without bound as one includes more of the degrees of freedom. The proof reveals the underlying mechanism: The spectral criterion that guarantees an emergent type III factor forces the logarithms of the Schmidt weights of the underlying state to develop unbounded fluctuations through a divergent concentration function. We then show that the inverse of the concentration function upper-bounds the overlap of the vector with any vector that has a flat spectrum, even after arbitrary basis changes on the two sides of the bipartition. Since vectors describing bipartite stabilizer states have flat Schmidt spectra on their support, their maximal fidelity with the state therefore vanishes asymptotically, which in turn implies a divergent min-relative entropy of magic. This connects the emergence of a type III factor from increasingly non-flat spectral data with divergent magic through a common mechanism.

\subsection{Further evidence for the proposal}
We now want to elaborate on the reasons underlying the discussion from the perspective of quantum field theory. An introduction to basic facts about operator algebras can be found in Appendix \ref{app:operator-algebra-intro}. In
\emph{algebraic quantum field theory} (AQFT) \cite{HaagKastler1964,Haag:1996hvx}
the full theory is described by a net of $C^\ast$-algebras, each algebra representing a causally complete region, and a state $\w$ on that algebra, which represents the vacuum.
A Hilbert space can then be obtained from the \emph{Gelfand–Naimark–Segal} (GNS) construction \ref{app:GNS}. This Hilbert space contains vectors that represent excitations that arise from $\w$ by inserting operators from the algebra. In this Hilbert space one can then complete the $C^\ast$ algebra associated to a causally complete region 
$D$ to a \emph{von Neumann algebra} $M_D$. Under the
assumption of asymptotic conformal invariance, relativistic covariance of the
net forces \cite{Buchholz:1986bg,Fredenhagen1985} that these algebras are hyperfinite type III$_1$ von Neumann algebras (see, e.g., Refs.~\cite{Haag:1996hvx, Witten:2018zxz, Sorce:2023fdx} for more pedagogical treatments of these insights).
One of the features of type III algebras is that they do not admit a \emph{normal tracial state} \cite{Sunder},
i.e., a map $\w:\cA \rightarrow \mathbb{C}$ with
\begin{equation}
    \w(AB) = \w(BA) < \infty,\quad A,B \in \cA.
\end{equation}
However, if the state locally has the form~\eqref{eq:projection_state}, then it
is easy to see that it is tracial for every local operator that maps the subspace associated to $P$ into itself. This is our main
point of view on why there is a fundamental tension between having a stabilizer
system and a local type III algebra. Another reason is 
that one can explicitly
compute R\'enyi entropies 
of vacuum states of CFTs \cite{Calabrese:2009qy}, which for single intervals follow the
universal formula 
\begin{equation}
    S_n(L) = \frac{1}{1-n} \log \Tr \r^{n} = \frac{c+\bar{c}}{12}(1 +
    \frac{1}{n})\log(\frac{L}{\e}) + c',
\end{equation}
where $c,\bar{c}$ are the chiral- and anti-chiral central charges, $c'$ is a non-universal constant, and $\e$ a UV cutoff. The fact that the R\'enyi entropy
depends on the index $n$ is only possible if the underlying spectrum is
non-flat. Another argument for the non-flatness of the entanglement spectrum is
the \emph{Unruh effect} \cite{Unruh:1976db}, which establishes that the vacuum of
Minkowski spacetime appears like a thermal state to a \emph{Rindler observer} that
experiences a constant acceleration.
This suggests that the reduced density matrix 
of the vacuum is
\begin{equation}
    \r_R = \frac{e^{-2 \pi K_R}}{Z},
\end{equation}
where
\begin{equation}
    K_R = \int_{x_1>0} \text{d}^{d-1}x\, x_1\, T_{00},
\end{equation}
is the restriction of the boost generator to the region causally accessible to
this observer. All these expressions have to be taken with care, since due to
the type III$_1$ nature of the local algebras, the reduced density matrix is
not rigorously defined. The algebraic realization of the Unruh effect is the \emph{Bisognano-Wichmann
theorem} \cite{Bisognano:1976za}, which states that the modular operator for the local algebra $\cA_W$ associated to a wedge $W$
is given by
\begin{equation}
    \D = e^{-2\pi K},
\end{equation}
where $K$ is the full boost operator. The boost operator has a spectrum over
the full line, which is impossible if $\r$ is flat. More generally, it has been
shown by Connes \cite{Connes:1971,Connes:1973} that for any type III$_1$
factor $\cA$ one has
\begin{equation}
    \G(\cA) = \bigcap_{\y \in \cA_*} \mathrm{spec}(\D_\y) = \mathbb{R}_+,
\end{equation}
where the intersection is over all \emph{faithful normal} states of $\cA$,
showing that the modular spectrum has nontrivial structure intrinsic to the type III$_1$ factor. This is consistent with, but does not by itself establish, the finite-cutoff picture of non-flat reduced spectra.
All these considerations suggest that our fundamental idea is
consistent with relativistic invariance.

The remainder of this paper is structured as follows: In Section \ref{sec:main-thm} we briefly introduce the framework and our main result as Corollary \ref{thm:no_type_III} which excludes the appearance of type III algebras in stabilizer systems. The results of this section are proven in Appendices \ref{app:lattice_systems} and \ref{app:notypeIII}. In Section \ref{sec:short-range-thm} we give an explicit decomposition of the local algebra for systems with short-range order, thus providing an example with total control over the state and the structure of the algebras. The associated proofs are given in Appendix \ref{app:short_range}. We contrast this in Section \ref{sec:long-range-order}, by giving an example which does not allow for a simple decomposition due to long-range order in the system. Finally, in Section \ref{sec:infinite-magic-implies-type-III}, we prove that for the Araki-Woods factors, the emergence of type III factors is equivalent to the divergence of the locally optimized min-relative entropy of magic \cite{Liu:2020yso}, thereby providing the first example that rigorously establishes a direct implication from infinite magic to the existence of type III factors. The proof reduces the optimized stabilizer fidelity to a comparison between the Schmidt spectrum and flat probability distributions, and then uses anti-concentration of the logarithmic Schmidt weights to show that this comparison vanishes precisely in the type III regime. The associated proofs are given in Appendix \ref{app:araki-woods}. We end the paper in Section \ref{sec:discussion} with a discussion of the results.

\subsection{Relation to recent work} 

As mentioned above, the idea that magic is a necessary ingredient for the emergence of a type III algebra from lattice systems was put forward by some of the present authors in Ref.~\cite{Chemissany:2025vye}. During the preparation of this manuscript, Refs. \cite{Benedetti:2026mfy} and \cite{Moosa:2026nqz} appeared, which improve on this proposal and provide more rigorous descriptions. Specifically, the work \cite{Benedetti:2026mfy} phrases this idea under the perspective of renormalization group flows
and proposes that a quantum system defined by a stabilizer system cannot flow to a relativistic quantum field theory in the infrared due to its flat spectrum. Their discussion strongly overlaps with the one of this introduction.
Our work provides a concrete framework in which these proposals can be studied explicitly.
In Ref.~\cite{Moosa:2026nqz} the author proposes a related framework analogous to the setting of Ref.~\cite{Chemissany:2025vye} that could also be adapted to the present setting. There, a sequence 
of lattice systems is considered and the algebras emerging in the inductive limit are analyzed. It is proven that, if the min-relative entropy of magic stays bounded as the limit is taken, the inductive limit of local algebras contains finite projections which implies that the algebras cannot purely be type III. This is a direct formalization of our previous statement that a finite injection of magic should not give rise to type III algebras.
Our work directly extends this result in two aspects.
The existence of a finite projection itself only implies that the algebra contains a finite corner, not that every factor in the central decomposition contains a finite projection. Our Corollary~\ref{thm:no_type_III} gives a stronger result for pure stabilizer systems, i.e., systems with zero magic. We establish that every factor in the central decomposition of the local algebras of the infinite system is semifinite and thus no type III factors can appear in the central decomposition. Furthermore \cite{Moosa:2026nqz} does not prove an inverse statement, namely whether infinite magic implies the emergence of a type III algebra. We provide a first result in this direction by Theorem \ref{thm:aw_magic}. We prove that for the explicit case of the hyperfinite factors studied by Araki and Woods \cite{ArakiWoods1968} the algebra is type III if and only if the underlying sequence of vectors has divergent locally optimized min-relative entropy of magic.

% Results begin here
\section{Results}
\label{sec:results}
We will now state our framework and main results. It should be noted that related results are available in the 
mathematical literature on topological phases \cite{Ogata:2022qdm,Jones:2023xew}, which analyze topological systems. In particular \cite{Ogata:2022qdm} proved that algebras associated to cone-like regions in the $\mathbb{Z}_2$ quantum double model on an infinite lattice also known as the surface code -- which is a stabilizer system -- are not of type III. This can be considered a special instance of our theory below which treats general stabilizer systems on infinite lattices. Our proof that establishes the existence of finite projections from a limit of the projections associated to finite regions was directly inspired by the proof technique used by 
Ref.~\cite{Ogata:2022qdm} for the toric code. An introduction to operator algebras can be found in Appendix \ref{app:operator-algebra-intro}, which introduces all technical terms used 
below.

\subsection{No type III algebras in stabilizer systems}\label{sec:main-thm}

In this section, we introduce our setup, establish much of the formal framework, and state our main result: Given an infinite stabilizer system
and a subregion, the associated local von Neumann algebra is not of type III. This will be established by proving that local algebras in stabilizer systems are decomposed into factors which contain so-called finite projections. For an introduction to the type classification and the relevance of finite projections in it, see Appendix \ref{app:type-recap}. 
\subsubsection{The setup}
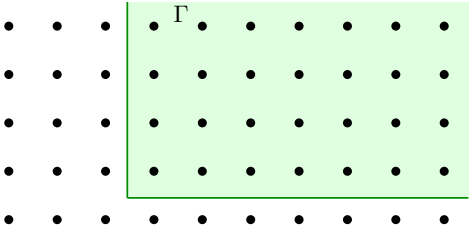
\begin{figure}[ht]
    \begin{tikzpicture}[x=.64cm,y=.64cm,
    site/.style={circle,fill=black,inner sep=1.2pt},
    support/.style={line width=1.1pt}]
  \fill[green!12] (2.45,.45) rectangle (9.5,4.5);
  \draw[green!55!black,line width=.65pt] (2.45,.45) -- (9.5,.45);
  \draw[green!55!black,line width=.65pt] (2.45,.45) -- (2.45,4.5);
  \foreach \x in {0,...,9}
    \foreach \y in {0,...,4}
      \node[site] at (\x,\y) {};
  % Both stabilizers act by X at their common exterior site, but their
  % interior supports are disjoint.
   \node[black,anchor=east] at (3.9,4.25) {$\G$};
\end{tikzpicture}%
    \caption{Illustration of a region $\G$ contained in a lattice.}
    \label{fig:subregion}
\end{figure}

We will
here only shortly define the relevant objects, with details relegated to Appendix
\ref{app:lattice_systems}. The content of this subsection will be presupposed in the following and referred to as \emph{the setup}.
On a lattice $\cL$  with a qubit Hilbert space at each lattice point, we consider the $C^\ast$-algebra of quasi-local operators
\begin{equation}
    \cA = \overline{\bigcup_{\G \Subset \cL} \cA_\G}^{\norm{.}},
\end{equation}
where we write $\G \Subset \cL$ to denote a finite subregion of $\cL$, with the whole expression forming the norm-completion of the union of operators supported on finitely many lattice sites. A consistent stabilizer system $\cS$ is a commuting set of Pauli operators for which finite products do not contain $-1$. 
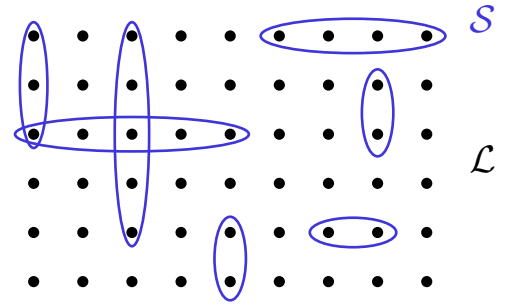
\begin{figure}[ht]
\begin{tikzpicture}[x=0.65cm,y=0.65cm]
\definecolor{stabilizerContourBlue}{HTML}{4035D8}
\tikzset{
  stabilizer site/.style={circle,fill=black,draw=none,
    inner sep=0pt,minimum size=4.2pt},
  stabilizer support/.style={draw=stabilizerContourBlue,line width=1pt,
    rounded corners=4pt,line cap=round,line join=round}
}
  \hspace{0.3cm}
  \providecommand{\cL}{{\mathcal L}}
  \providecommand{\cS}{{\mathcal S}}
  % Blue contours follow the arrangement in the supplied sketch.
  \draw[stabilizer support] (0,4)
    ellipse[x radius=0.182cm,y radius=0.832cm];
  \draw[stabilizer support] (2,3)
    ellipse[x radius=0.2275cm,y radius=1.482cm];
  \draw[stabilizer support] (2,3)
    ellipse[x radius=1.547cm,y radius=0.2275cm];
  \draw[stabilizer support] (6.5,5)
    ellipse[x radius=1.222cm,y radius=0.2275cm];
  \draw[stabilizer support] (7,3.435)
    ellipse[x radius=0.208cm,y radius=0.57525cm];
  \draw[stabilizer support] (4,0.47)
    ellipse[x radius=0.208cm,y radius=0.5525cm];
  \draw[stabilizer support] (6.5,1)
    ellipse[x radius=0.572cm,y radius=0.195cm];
  \foreach \x in {0,...,8}{
    \foreach \y in {0,...,5}{
      \node[stabilizer site] at (\x,\y) {};
    }
  }
  \node[stabilizerContourBlue,anchor=west,font=\Large] at (8.70,5.35) {$\cS$};
  \node[anchor=west,font=\Large] at (8.70,2.5) {$\cL$};
\end{tikzpicture}
\caption{Visualization of a stabilizer system. The support of individual stabilizers is indicated by blue ellipses.}
\end{figure}
Given a consistent stabilizer system, for any finite region $F \Subset \cL$, the stabilizers define a state that is given by the projection onto the code subspace of the local stabilizer group. We prove in Appendix \ref{app:lattice_systems} as Proposition \ref{prop:define_state} that 
\begin{theorem}[Stabilizer systems induce a canonical state]
    Let $\cL$ be a lattice and $\cS$ a consistent set of stabilizers. Then there exists a unique state $\w$ on the quasi-local $C^\ast$-algebra $\cA$ which has reduced density matrices 
    \begin{equation}
        \rho_F = \frac{P_F}{\operatorname{Tr}P_F},
    \end{equation}
    where 
    \begin{equation}
        P_F = \frac{1}{|\cG^{\cS}_F|}\sum_{g \in \cG^{\cS}_F} g
    \end{equation}
    is the projection onto the local code subspace defined by $\cG^{\cS}_F$, the local stabilizer group defined in Eq.~\eqref{eq:local_stabilizer_group}. 
\end{theorem}
This state induces a GNS Hilbert
space $\cH_\w$ \cite{Blackadar} on which we have a representation $\pi_\w$ of $\cA$. For a given
subregion $\G \subset \cL$, we similarly define the quasilocal algebra
$\cA_\G$ via
\begin{equation}
    \cA_\G = \overline{\bigcup_{\L \Subset \G} \cA_\L}^{\norm{.}},
\end{equation}
An example of such a region is given in Figure \ref{fig:subregion}.

For this algebra, its associated von Neumann algebra is given by
\begin{equation}
    M_{\G}:=\pi_\omega\!\left(\cA_\G\right)'',
\end{equation}
where for a set $S \subset \cB(\cH)$ of operators, the commutant $S'$ is the set of operators that commute with all operators in $S$. Here note that we needed to first represent the abstract algebra $\cA_\G$ on the Hilbert space $\cH_\w$. The abstract algebra $\cA_\G$ has no associated Hilbert space and has to be distinguished from its concrete representation $\cU_\G := \pi_\w(\cA_\G)$.
Its associated von Neumann algebra $M_\G$ describes the degrees of freedom that create excitations of the ``vacuum'' $\w$ that are located in $\G$. This is our main object of study which we aim to characterize.
\subsubsection{The classification theorem}
We are now ready to state our main theorem, but first we want to comment on its content.
In standard quantum mechanics, one considers algebras such as 
\begin{equation}
    M_A = \cB(\cH_A) \otimes \id,
\end{equation}
for some bipartite system $\cH = \cH_A \otimes \cH_B$. Such an algebra is called a factor for obvious reasons. More generally an algebra is called a factor if its center  $\cZ(M) = M \cap M'$ is trivial, i.e., $\cZ(M) = \bbC \id$. This does not cover all algebras. For example, every type I von Neumann algebra $M$ on a finite-dimensional Hilbert space $\cH$ induces a basis such that 
\begin{equation}
    \cH = \bigoplus_\a \cH^{\a}_A \otimes \cH^{\a}_B,
\end{equation}
and the algebra $M$ in this basis is 
\begin{equation}\label{eq:central-decomp}
    M = \bigoplus_\a \cB(\cH^{\a}_A) \otimes \id^\a_B.
\end{equation}
We see that the algebra becomes a direct sum of factors. 
The center is then generated by the operators
\begin{equation}
    z_\w =  \left( \bigoplus_{\a \neq \w} 0 \right) \oplus \id_\w,
\end{equation}
i.e., operators that  act nontrivially on a single sector. These projections are called \emph{central projections}.
For each such $z_\w$ we can consider the corner $z_\w M$, which describes $M$ only on the sector $\w$, i.e., 
\begin{equation}
   z_\w M  =  \left( \bigoplus_{\a \neq \w} 0 \right) \oplus \left( \cB(\cH^\w_A ) \otimes \id^\w_B \right).
\end{equation}
This decomposition generalizes and one can represent any von Neumann algebra $M$ as 
\begin{equation}
    M = \int_X \mathrm{d} x\ M_x,
\end{equation}
with an appropriate notion of integrals over some measure space $X$. The central projections then become associated to subsets $\w \subset X$
\begin{equation}
    z_\w = \int_X \mathrm{d}x\ \chi_\w(x),
\end{equation}
where $\chi_\w$ is a characteristic function
\begin{equation}
    \chi_\w(x) = \begin{cases}
        1,\ x \in \w, \\
        0,\ x \notin \w.
    \end{cases}
\end{equation}
The main point is that the algebras $M_x$ are also factors. We see that a von Neumann algebra can generally be considered as being built as a direct sum of its factors,
and since the factors act diagonally on orthogonal subspaces described by central projections $z_\w$, they concern independent degrees of freedom.
In Ref.~\cite{Moosa:2026nqz} it has been proven that having bounded magic
implies that at least one of the factors $M_x$ is not type III. Thus, the full algebra is not type III. This was established by demonstrating that there exists a projection $q\in M$ that is finite. This means that there exists a state on the corner $q M q$ which is tracial, i.e., there is a $\w : qMq \to \bbC$ so that 
\begin{equation}
\w(AB) = \w(BA) < \infty,\quad A,B \in qMq.
\end{equation}
Here it is important that the value is finite, i.e., also $\w(q)< \infty$. 
The operator $q$ can similarly be decomposed 
\begin{equation}
    q = \int_X \mathrm{d}x\ q_x.
\end{equation}
This allows for $q_x = 0$ for some $x$, so that the corner $q M q$ loses information about the algebra $M$ and only considers the parts where $q_x \neq 0$. The theorem of Ref.~\cite{Moosa:2026nqz} does not make any statements about the rest of $M$, which is annihilated by $q$. Here we establish a stronger claim for stabilizer systems directly, namely
\begin{proposition}[Finite projections in every central sector]
    Let $M_\G$ be the von Neumann algebra associated to a subregion $\G$ in the above setup, and let $z \in \cZ(M_\G)$ be a nonzero central projection. Then $zM_\G$ contains a nonzero finite projection.
\end{proposition}
This is proven as Theorem \ref{thm:finite_projections_exist} in Appendix \ref{app:notypeIII}.
The non-existence of finite projections is a defining feature of type III algebras \cite[III.1.4.2]{Blackadar}, which thus implies
\begin{corollary}[Exclusion of type III central summands]\label{thm:no_type_III}
    For every nonzero $z \in \cZ(M_\G)$, the associated subalgebra $zM_\G$ is not of type III.
\end{corollary}
We believe this is the most general statement one can make within this setting to exclude the emergence of a quantum
field theory in the infinitely large lattice with the technology currently available. As mentioned previously, 
without a more rigorous formulation of effective field theory which explicitly realizes the EFT via subalgebras of local algebras within the lattice
this is still not a complete argument. In the systems we consider, the lattice spacing is kept fixed and we only assume that the lattice is infinitely large. One could also consider sequences of lattices with an explicit lattice spacing that vanishes in some limit to reach a continuum description.

\subsection{Algebras in stabilizer systems without long-range order}\label{sec:short-range-thm}
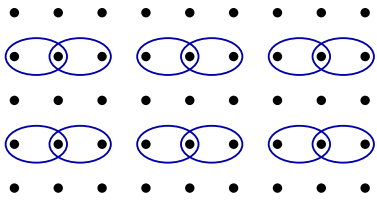
\begin{figure}[ht]
\begin{tikzpicture}[x=0.58cm,y=0.58cm,
    site/.style={circle,fill=black,inner sep=1.25pt},
    support/.style={draw=blue!65!black,line width=.7pt}]
  \foreach \x in {0,...,8}
    \foreach \y in {0,...,4}
      \node[site] at (\x,\y) {};
  % Each unfilled loop is the boundary of a bounded stabilizer support.
  \draw[support] (.5,3) ellipse [x radius=.7,y radius=.42];
  \draw[support] (1.5,3) ellipse [x radius=.7,y radius=.42];
  \draw[support] (3.5,3) ellipse [x radius=.7,y radius=.42];
  \draw[support] (4.5,3) ellipse [x radius=.7,y radius=.42];
  \draw[support] (6.5,3) ellipse [x radius=.7,y radius=.42];
  \draw[support] (7.5,3) ellipse [x radius=.7,y radius=.42];
  \draw[support] (.5,1) ellipse [x radius=.7,y radius=.42];
  \draw[support] (1.5,1) ellipse [x radius=.7,y radius=.42];
  \draw[support] (3.5,1) ellipse [x radius=.7,y radius=.42];
  \draw[support] (4.5,1) ellipse [x radius=.7,y radius=.42];
  \draw[support] (6.5,1) ellipse [x radius=.7,y radius=.42];
  \draw[support] (7.5,1) ellipse [x radius=.7,y radius=.42];
\end{tikzpicture}
\caption{A stabilizer system that induces short-range order as it consists of disconnected clusters. The support of individual stabilizers is indicated by ellipses.}
\label{fig:stabilizers-short-range-order}
\end{figure}

The previous section provides a structure theorem for subalgebras in stabilizer systems. This raises the question of whether we can obtain more detailed information on the structure of the algebras.
We now discuss our theory for the simpler case of a version of \emph{short-range ordered}
states. As discussed more deeply in Appendix \ref{app:short_range}, we say that a stabilizer system has short-range order in the region $\G$ if the associated state
is made up of products of states living in finite-sized regions. An example of short-range order induced by a stabilizer system is illustrated in Figure \ref{fig:stabilizers-short-range-order}. There, the stabilizers appear in disjoint clusters, which leads to a stabilizer group that is a tensor product of local stabilizer groups and thus the state defined by the code space projections factorizes into a tensor product between these clusters.
In a more conventional language, this can be understood as the existence of a \emph{local} Clifford circuit which provides a basis transformation that diagonalizes $\r_F$ \cite{NielsenChuang2010}.
This implies that there is a basis choice for which the reduced density matrix can be written as
\begin{equation}
    \rho_{F} =
 \ket{\mathbf 0}\!\bra{\mathbf 0}
 \otimes\frac{\id}{D},
\end{equation}
where the vectors $\ket{0}$ arise from degrees of freedom that are fixed by the stabilizers and the identity component arises from either stabilizers crossing the boundary $\partial\G$ or degrees of freedom in the code space, which are not fixed by the stabilizers. This is well known in the quantum information theory literature. Our contribution is to extend this to provide a precise meaning of this in an infinite lattice.
The key point is that we can choose these bases consistently such that the state can be thought of as
\begin{equation}\label{eq:growing_state}
    \r_{\G'} \cong \r_\G \otimes \r'_{\d \G},
\end{equation}
so that the state on $\G$ is an infinite tensor product.
The full algebra $M_\G$ then becomes the tensor product of an algebra acting on
the pure, internally fixed modes and an algebra on the maximally mixed modes
that arise either from cross-boundary stabilizers or the projection onto the
code subspace, i.e.,
\begin{equation}
    M_\G \cong \cB(\cH) \otimes \cR.
\end{equation}
Its type is determined by whether each collection contains finitely or
infinitely many modes.  Finitely many pure modes give a finite-dimensional type-I factor,
whereas infinitely many modes give a type-I$_\infty$ factor, since $\cH$ is then
either finite- or infinite-dimensional.  Finitely many maximally
mixed modes give again a finite-dimensional type-I factor, while infinitely many produce the hyperfinite
type-II$_1$ factor. Their tensor product is consequently of type I or type II,
and never of type III. 
In this case we obtain Theorem \ref{thm:short_range_algebra}, which we informally present as
\begin{theorem}[Short-range stabilizer algebras (informal)]\label{thm:type_for_short_range_entanglement}
    Let $\cS$ be a consistent stabilizer system on a lattice $\cL$ and $\G$ a subregion on which $\cS$ induces short-range order. Then the von Neumann algebra $M_\G$ associated to this region in the state $\w$ is isomorphic to a tensor product of hyperfinite factors of type I or type II.
\end{theorem}
\subsection{Long-range order and central sectors}\label{sec:long-range-order}
The preceding argument does not apply when long-range order imprints further
structure.  For example, consider the chain
$\cL=\mathbb Z$ with stabilizers
\begin{equation}
 S_n=Z_nZ_{n+1}.
\end{equation}
This system is visualized in Figure \ref{fig:long-range-order}:
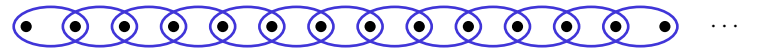
\begin{figure}[ht]
    \centering
    \begin{tikzpicture}[x=0.65cm,y=0.65cm]
\definecolor{stabilizerContourBlue}{HTML}{4035D8}
\tikzset{
  stabilizer site/.style={circle,fill=black,draw=none,
    inner sep=0pt,minimum size=4.2pt},
  stabilizer support/.style={draw=stabilizerContourBlue,line width=1pt,
    rounded corners=4pt,line cap=round,line join=round}
}
  \foreach \n in {0,...,12}{
    \draw[stabilizer support] ({\n+0.5},0)
      ellipse[x radius=0.49cm,y radius=0.26cm];
  }
  \foreach \n in {0,...,13}{
    \node[stabilizer site] at (\n,0) {};
  }
  \node[black,anchor=west] at (13.75,0) {$\cdots$};
\end{tikzpicture}
\caption{Visualization of a stabilizer system inducing long-range order.}
    \label{fig:long-range-order}
\end{figure}

For a finite interval $\Gamma$, the canonical local state is proportional to
\begin{equation}
    \rho_\Gamma=\frac12\bigl(\ket{\mathbf0}\!\bra{\mathbf0} +\ket{\mathbf1}\!\bra{\mathbf1}\bigr).
\end{equation}
The two terms label the all-zero and all-one sectors. The Hilbert space takes the form 
\begin{equation}
    \cH = \cH_0 \oplus \cH_1,
\end{equation}
where $\cH_0$ describes excitations over the vector $\ket{0}$ and $\cH_1$ analogously for the vector $\ket{1}$.
For the half-chain
indexed by positive labels $n>0$, the local algebra $\cM_>$ thus becomes
\begin{equation}
    \cM_> = \big( \cM^{0}_> \otimes \id_< \big) \oplus \big( \cM^{1}_> \otimes
        \id_< \big) ,
\end{equation}
where $\cM^{0/1}_>$ is a type I$_\infty$ factor $\cB(\cH_>)$. The all $0$ vector has
long-range order, in the sense that the strong limit
\begin{equation}
    Z_n \xrightarrow{\mathrm{s}} P
\end{equation}
exists and is nontrivial as $n \to \infty$. $P$ is an operator that distinguishes the two sectors and is central for the local algebra
$M_\G$. By this we mean that for all vectors $\ket{\y} \in \cH_0$ one has that 
\begin{equation}
    \lim_{n \to \infty} Z_n \ket{\y} = \ket{\y},
\end{equation}
because all vectors in $\cH_0$ have qubits asymptotically in the vector $\ket{0}$, whereas for all vectors $\ket{\p}\in \cH_1$ one has 
\begin{equation}
    \lim_{n \to \infty} Z_n \ket{\p} = - \ket{\p}.
\end{equation}
Thus $\cH_0$ is the $+1$ eigenspace of the operator $P$, whereas $\cH_1$ is the $-1$ eigenspace of $P$. $P$ is central for the local algebras, because $Z_n$ will commute asymptotically with every local operator. The existence of a nontrivial center is one of the defining features of long-range order, see e.g. \cite{LanfordRuelle1969} for the relation to short-range correlations. In this case, we still have type I factors appearing in the central
decomposition. Due to the emergence of a center from observables ``at infinity'', we cannot use the same argument as for the proof of Theorem \ref{thm:type_for_short_range_entanglement}. To prove this more generally, we need control over these
observables, how to find them for general stabilizer systems and then potentially to prove the type for each individual factor appearing in the central decomposition, which is outside the scope of this work. One way to identify these sectors could use the framework developed in Ref.~\cite{Haah:2013oba} to recover superselection sectors from local data.  
%\AJ{What are these citations for?}\LS{Added some explanation, removed ogatas.}

The same phenomenon can lead to an uncountable number of sectors.  On $\mathbb Z^2$,
take vertical bond stabilizers
\begin{equation}
 S_{i,n}=Z_{i,n}Z_{i,n+1}.
\end{equation}
Fixing a sector now requires one binary choice for every vertical line, hence
$2^{\aleph_0}$ choices. The expected decomposition of the algebra associated
to a single vertical region is then a direct integral over this continuum of
choices, rather than the finite direct sum of the one-dimensional example.
 
\subsection{Type III algebras and infinite magic}\label{sec:infinite-magic-implies-type-III}
The discussion of the previous sections alone only establishes that systems without magic, which arise from stabilizers, do not contain type III factors. The action of a unitary operator supported on a finite region will not change the underlying Hilbert space and therefore does not change the type of the local algebras. This suggests that one indeed needs to change the state on an infinite number of degrees of freedom to produce a type III factor. We would like to argue that this infinite change has to be accompanied with infinite magic. We can prove this for a simple toy model:
We consider the standard bipartite Araki--Woods factors, i.e., \emph{infinite tensor products of type I} (ITPFI) factors.
These are factors with vacua of the form
\begin{equation}
  \ket{\psi^N}=\bigotimes_{i=1}^{N}\ket{\psi_i},
  \qquad
  \ket{\psi_i}\in\bbC^2_{A,i}\otimes\bbC^2_{B,i},
\end{equation}
and the limiting algebra on the $A$ degrees of freedom is obtained as the
inductive limit of the algebras acting on finitely many of the spaces labeled
with $A$, see e.g.\ Refs.~\cite{vonNeumann1939,ArakiWoods1968,Witten:2018zxz} for more details. In this setting we can prove that infinite magic
is necessary and sufficient for a type III factor. We will henceforth call the index $N$ the \emph{layer index} and say that $\psi^N$ is a state at level $N$.
For a state at level $N$, 
the \emph{stabilizer fidelity} \cite{Bravyi:2018ugg} is defined by
\begin{equation}\label{eq:stabilizer_fidelity}
    F_{\mathrm{STAB}}^{(N)}(\psi^N)
 := \sup_{\phi\in\mathrm{STAB}_{2N}}\
 \left|\braket{\phi}{\psi^N}\right|^2 , 
\end{equation}
where $\Stab_{2N}$ denotes
the stabilizer states on the $2N$ qubits.
In analogy, we define the \emph{locally optimized stabilizer fidelity}
\begin{equation}\label{eq:locally_optimized_stabilizer_fidelity}
    F_{\mathrm{STAB},\mathrm{LO}}^{(N)}(\psi^N)
 := \sup_{U_A,U_B\in\cU(\bbC^{2^N})}\ F_{\mathrm{STAB}}^{(N)}\big( (U_A\otimes U_B) \psi^N \big) .
\end{equation}
To this we associate a locally optimized version of the \emph{min-relative entropy of magic} \cite{Liu:2020yso}
\begin{equation}\label{eq:locally_optimized_min_relative_entropy}
    \cD_{\mathrm{min},\mathrm{LO}}^{(N)}(\psi^N)
 :=-\log F_{\mathrm{STAB},\mathrm{LO}}^{(N)}(\psi^N).
\end{equation}
We then have the following result.

\begin{theorem}[Type III factors and divergent magic]\label{thm:aw_magic}
Let $\ket{\psi^N}=\bigotimes_{i=1}^{N}\ket{\psi_i}$ be a sequence of vectors
that define the vacuum of an infinite tensor product, 
and let $M_A$ be the limiting factor
generated on the $A$ degrees of freedom.  Then
\begin{itemize}
 \item $M_A$ is of type I or type II if and only if
 \begin{equation}
     \sup_{N}\cD_{\mathrm{min},\mathrm{LO}}^{(N)}(\psi^N)<\infty.
 \end{equation}
 \item $M_A$ is of type III if and only if
\begin{equation}
    \sup_{N}\cD_{\mathrm{min},\mathrm{LO}}^{(N)}(\psi^N)=\infty.
\end{equation}
\end{itemize}
\end{theorem}

The proof of this theorem is given in Appendix~\ref{app:araki-woods}.  It demonstrates that,
for the restricted model of Araki-Woods factors, infinite magic is indeed
necessary and sufficient for the emergence of a type III factor. 
A natural next question is whether there is a qubit-independent notion
of nonstabilizerness appropriate for more general settings.

\section{Discussion} \label{sec:discussion}

This %paper 
work
establishes a rigorous 
and quantitative relationship between the flatness of the spectrum arising from stabilizer systems and the impossibility to have local algebras
that coincide with the ones of a relativistic quantum field theory. This establishes the proposals of Refs.~\cite{Chemissany:2025vye, Benedetti:2026mfy} in a concrete framework and extends the results of Ref.~\cite{Moosa:2026nqz}.
In particular this provides a rigorous perspective on recent findings of divergent magic measures
at critical points of quantum systems
\cite{oliviero2022magicstateresourcetheory} and the relevance of magic in the emergence of spacetime \cite{cao2024gravitationalbackreactionmagical, Cao:2026uoq} as we could demonstrate that, for the case of Araki-Woods factors, infinite magic is equivalent to the emergence of a type III factor.
This provides strong evidence for the incompatibility of stabilizer codes and the properties of quantum field theories. This suggests that the conventional 
holographic quantum error-correcting codes \cite{Pastawski:2015qua} based on stabilizer quantum error correction are fundamentally unsuitable as a model of boundary conformal field theories for holography. 
In particular, we proved that in the central decomposition of local algebras arising from stabilizer systems, no factors of type III can appear. 
Additionally, we gave a complete characterization of the local algebras for
stabilizer systems that induce short-range order in a region in terms of tensor products of
hyperfinite factors of type I or type II. At last we proved that in the case of the Araki-Woods factors, a type III algebra can only arise if the underlying sequence of states has infinite magic.

While we present a comprehensive answer in this work, several compelling %questions 
directions remain open:
It would be interesting to understand if one can extend our techniques and the approach of Ref.~\cite{Moosa:2026nqz}
to determine when unbounded nonlocal magic implies the existence of a type III algebra more generally.
In particular, our proof shows that the logarithm of the probability
distribution of the bipartite spectrum develops infinite fluctuations, which
demonstrates the emergence of a continuum in the 
spectrum of the modular operator. It remains unclear whether %if
one can
understand this more deeply, for example by analyzing which properties of circuits
lead to such a spectrum and whether one can prove the non-existence of finite
projections in this case. Another related study performed in \cite{Sahu:2026xyt} found type II$_1$ algebras in random tensor networks in the infinite bond dimension limit. It is unclear whether this arises from a similar mechanism as we studied here, since random tensor networks have flat spectra at leading order in the inverse bond dimension \cite{Hayden:2016cfa}. A similar construction appeared in \cite{Soni:2023fke}, which recovered an algebraic structure reminiscent of the crossed product associated to subalgebras in perturbative quantum gravity \cite{Witten:2021unn,Chandrasekaran:2022cip,Chandrasekaran:2022eqq,Jensen:2023yxy,Kudler-Flam:2023qfl} that gives rise to type II$_\infty$ algebras associated to a subregion. Since gravity is a gauge theory and imposing gauge invariance on the states forces edge modes at the interface of two regions to form maximally entangled states \cite{Dong:2023kyr}, it would be interesting to understand whether the reason for the emergence of type II$_\infty$ factors in gravity is related to this maximal entanglement, as this is the mechanism behind the emergence of type II$_\infty$ factors in the short-range ordered case above. One drawback of our approach is that we keep the lattice fixed and have no explicit length scale that we can use to formulate a continuum limit. One natural extension is to consider sequences of stabilizer systems which have a length scale and understand whether one can make similar claims in 
this setup. For example in Refs.~\cite{Osborne:2019bsq, Stottmeister:2020ezd, Morinelli:2020uea, Osborne:2021ppp, Stottmeister:2022ptp, Stottmeister:2023wza} continuum limits of different lattice systems were studied and the convergence to quantum field theories could be rigorously established. One extension of our work would be to translate the setup to such systems and show that no type III algebras emerge there either. Similarly, the scaling algebra approach to the renormalization group in algebraic quantum field theory developed in Refs.~\cite{Buchholz:1995gr,Buchholz:1995mx,Verch:1996yf,Buchholz:1997vu,Bostelmann:2007ut,Bostelmann:2009uec} may provide a useful framework to which to translate our approach. In particular, since the presence of Lorentz symmetry and asymptotic scale invariance in relativistic quantum field theories underlie the universality of local type III algebras \cite{Fredenhagen1985}, it would be useful to understand whether one can connect the emergence of these continuum features to the presence of infinite magic.

On a broader level, this work helps delineate the quantum from the classical world by asking where a stabilizer description ceases to suffice. The boundary is subtle: Stabilizer systems can carry substantial entanglement, yet their local algebras cannot realize the type III structure characteristic of relativistic quantum field theory. From a complementary computational perspective, recent work shows that even deciding whether a general state lies in the stabilizer polytope can require super-exponential time in the number of qubits \cite{Leone:2026kxv}. Thus the resources that mark this boundary are not always easy to identify, even when their structural consequences are profound. We hope these results bring us closer to a question that is as old as it is compelling: What, precisely, must be present in a quantum state for the spacetime we inhabit to emerge?

\begin{acknowledgments}
We thank Elliott Gesteau for discussions. L.S., D.S.\ and A.J.\ acknowledge support by the Einstein Research Unit ``Perspectives of a quantum digital transformation'' as well as Berlin Quantum.
A.J.\ also acknowledges funding for his JRG Program position at the APCTP provided by the Science and Technology Promotion Fund and Lottery Fund of the Korean Government, as well as the Korean Local Governments of Gyeongsangbuk-do Province and Pohang City.
J.E.\ and L.B.\ have been supported by the BMFTR (MUNIQC-Atoms), the DFG (CRC 183, BoLaCo, and SPP 2514), the Quantum Flagship (Millenion, PasQuanS2), the Munich Quantum Valley, Berlin Quantum, QuantERA (SDPCode), and the European Research Council (DebuQC).\smallskip

\textbf{Note on AI usage:} The proofs of Appendix \ref{app:notypeIII} were constructed through close collaboration between the authors and GPT 5.4 and 5.5 Pro, which proposed incomplete first versions of the proofs that were subsequently completed by us together with GPT 5.5 Pro. This construction was only achieved after we suggested adjusting Ogata's technique from 
Ref.\ \cite{Ogata:2022qdm} to find a finite projection in our current setup. The notion of short-range order that led to the results in \ref{app:short_range} was discovered by us and an erroneous first version of the proof was produced by GPT 5.5 that was missing the equivalence between global and local representations that was subsequently worked out by us together with GPT 5.5 Pro. The proof of Theorem \ref{thm:aw_magic} was generated autonomously by GPT 5.5 Pro after we posed the question without any hint towards the solution. All of the theorem statements have been collaboratively formulated by the authors without involvement of AI.

\end{acknowledgments}
\onecolumngrid
\clearpage
\appendix
\setcounter{secnumdepth}{2}
\setcounter{tocdepth}{2}
\renewcommand{\thesubsection}{\thesection.\arabic{subsection}}
\renewcommand{\thesubsubsection}{\thesubsection.\alph{subsubsection}}
\makeatletter
\renewcommand{\p@subsection}{}
\renewcommand{\p@subsubsection}{}
\makeatother
\begin{center}
\textbf{\large Appendix}
\end{center}
\tableofcontents
\resumetoc
\newpage
\section{Operator algebraic preliminaries}\label{app:operator-algebra-intro}
This appendix collects some basic statements and fixes the notation on von Neumann algebras that we will use throughout this work.
    To define a von Neumann algebra, one needs to first introduce
    $C^\star$-algebras. A $C^\star$-algebra $\cC$ is a normed complex vector space together with a multiplication, and an involution $\dagger: a \rightarrow a^\dagger$.
    The involution is antilinear and satisfies 
    \begin{equation}
        a = (a^\dagger)^\dagger,\quad (ab)^\dagger = b^\dagger a^\dagger \ . 
    \end{equation}
    Additionally, in a $C^\ast$-algebra, the norm satisfies
    \begin{equation}
        \norm{a a^\dagger} = \norm{a}^2.
    \end{equation}
    By definition a $C^\ast$-algebra $\cA$ is complete in the norm-topology, i.e., if a sequence $a_n \in \cA$ converges in norm to $a$, then $a \in \cA$. This turns $C^\ast$-algebras into Banach spaces, i.e., normed vector spaces that are complete in the norm-topology. 
    The simplest example is the set of bounded linear operators $\cB(\cH)$ acting on a Hilbert
    space $\cH$. In general, a $C^\ast$- algebra can be defined without making an explicit reference to an underlying Hilbert space.
    In the remainder of this section, we fix a Hilbert space $\cH$ and will only consider unital
    algebras, i.e., those that contain the identity $\id$ on $\cH$. Defining a von Neumann algebra involves introducing the commutant $S'$ of a subset $S \subset \cB(\cH)$,
    which is defined as the set of bounded operators that commute with all of
    $S$, i.e.,
    \begin{equation}
        S' := \{ a \in \cB(\cH): [a,s] = 0\ \forall s \in S \}.\label{eq:define-commutant}
    \end{equation}
    A von Neumann algebra $M$ is then defined as a subalgebra of $\cB(\cH)$ that is closed under Hermitian conjugation and equal to its double commutant, i.e.,
    \begin{equation}
        M = M''.
    \end{equation}
    Now, given a self-adjoint subset $\cA \subset \cB(\cH)$, the double commutant
    \begin{equation}
        M := \cA''
    \end{equation}
    is always a von Neumann algebra, and is called the von Neumann algebra \emph{generated} by $ \cA$.
    The \emph{center} of a von Neumann algebra $M$ is the set
    \begin{equation}
        \cZ(M) = M \cap M' 
    \end{equation}
    Finally, a von Neumann algebra $M$ is called a factor if 
    \begin{equation}
        \cZ(M) = \mathbb{C} \id.
    \end{equation}
    A simple example of a factor can be found in a bipartite system whose Hilbert space $ \cH$ takes on the form
    \begin{equation}
        \cH = \cH_A \otimes \cH_B.
    \end{equation}
    Consider the algebra
    \begin{equation}
        \cA = \cB(\cH_A) \otimes \id,
    \end{equation}
    whose commutant reads
    \begin{equation}
        \cA' = \id \otimes \cB(\cH_B),
    \end{equation}
    and is therefore clearly a factor.
    \subsection{Topologies on $\cB(\cH)$}
    For the bounded operators on an infinite-dimensional Hilbert space $\cH$,
    there are several possible inequivalent topologies on them, which are
    crucial to the definition of von Neumann algebras. The best known example
    is the norm topology, which is the topology induced by the standard
    operator norm, i.e., for $A \in \cB(\cH)$,
    \begin{equation}
        \norm{A} = \sup_{\substack{\ket{\y}\in\cH\\\norm{\y}=1}} \norm{A\y}.
    \end{equation}
    This defines a topology, the norm topology, in which convergence is defined
    via
    \begin{equation}
        A_n \xrightarrow{\norm{\cdot}} A
        \quad\Longleftrightarrow\quad
        \norm{A_n-A}\to 0.
    \end{equation}
    This topology is very fine, as it is based on a uniform norm which takes
    the supremum over all vectors. There are weaker topologies: The
    $\sigma$-strong topology defines that a sequence of operators
    $A_n$ converges $\s$-strongly to an operator $A$, if for every sequence $(\ket{\y_k})_{k\in\mathbb{N}}$ in $\cH$ such that
    \begin{equation}
        \sum_{k=1}^{\infty}\norm{\y_k}^2<\infty,
    \end{equation}
    one has
    \begin{equation}
        \sum_{k=1}^{\infty}\norm{(A_n-A)\y_k}^2\to 0.
    \end{equation}
    This will be denoted by 
    \begin{equation}
        A_n \xrightarrow{\sigma\text{-}\mathrm{s}} A
    \end{equation}
    The strong topology is the simpler case where one has convergence for
    each single vector individually, i.e.,
    \begin{equation}
        A_n \xrightarrow{\mathrm{s}} A
        \quad\Longleftrightarrow\quad
        \norm{(A_n-A)\y}\to 0
        \quad\text{for all }\ket{\y}\in\cH.
    \end{equation}
    In the $\sigma$-weak (or ultraweak) topology a sequence $A_n$ converges to $A$, if for every two sequences $(\ket{\x_k})_{k\in\mathbb{N}}$ and $(\ket{\y_k})_{k\in\mathbb{N}}$ in $\cH$ satisfying
    \begin{equation}
        \sum_{k=1}^{\infty}\norm{\x_k}^2<\infty,
        \qquad
        \sum_{k=1}^{\infty}\norm{\y_k}^2<\infty,
    \end{equation}
    one has
    \begin{equation}
        A_n \xrightarrow{\sigma\text{-}\mathrm{w}} A
        \quad\Longleftrightarrow\quad
        \sum_{k=1}^{\infty}\mel{\x_k}{(A_n-A)}{\y_k}\to 0.
    \end{equation}
    Similarly, the weak topology is defined by
    \begin{equation}
        A_n \xrightarrow{\mathrm{w}} A
        \quad\Longleftrightarrow\quad
        \mel{\x}{(A_n-A)}{\y}\to 0
        \quad\text{for all }\ket{\x},\ket{\y}\in\cH.
    \end{equation}
    By von Neumann's bicommutant theorem \cite{TakesakiI,Sunder} von Neumann
    algebras are exactly the algebras which are closed in the weak topology. It
    also states that for these algebras, being weakly closed is equivalent to
    being closed in the strong topology. 
    The $\s$-weak topology arises from the notion of a normal linear functional:
\begin{definition}[Normal linear functional]
\thmref{cf.~III.2.1.2 in Ref.~\cite{Blackadar}}
Let $M$ be a von Neumann algebra. A bounded linear functional
$\P:M \rightarrow \mathbb{C}$ is called normal if, for every bounded increasing
net $(x_i)$ in $M_+$ with $x=\sup_i x_i$,
\begin{equation}
    \P(x) = \lim_i \P(x_i).
\end{equation}
\end{definition}
\begin{definition}[Predual]
\thmref{cf.~III.2, p.~261 in Ref.~\cite{Blackadar}}
Let $M$ be a von Neumann algebra. Let $M_*$ be the space of normal bounded linear
functionals on $M$. This is called the predual.
\end{definition}
The predual $M_\ast$ should be distinguished from the \emph{dual} $M^\ast$, which is the space of bounded linear functionals, without the demand of normality.
It has been discovered by Sakai, \cite{Sakai98} that von Neumann algebras can be
uniquely determined by their predual. In particular, von Neumann algebras are exactly the $C^\ast$-algebras that are the dual space of a Banach space, i.e., $M = E^\ast$ for some Banach space $E$. Thus, to characterize a von Neumann algebra, it is sufficient to characterize its predual. A useful representation of elements of the predual is given by:

\begin{theorem}[Normal functionals as sums of vector functionals]\label{thm:blackadar-normal-vector}
\thmref{cf.~III.2.1.4 in Ref.~\cite{Blackadar}}
Let $M$ be a von Neumann algebra on $\cH$ and let
$\varphi:M \rightarrow \mathbb{C}$ be a bounded linear functional. Then the
following are equivalent:
\begin{itemize}
    \item[(i)] $\varphi$ is normal.
    \item[(ii)] There exist $\psi_i,\eta_i \in \cH$ such that
    \begin{equation}
        \varphi(x) = \sum_i \mel{\psi_i}{x}{\eta_i},
        \quad x \in M,
    \end{equation}
    and
    \begin{equation}
        \sum_i \norm{\psi_i}^2 < \infty,
        \quad
        \sum_i \norm{\eta_i}^2 < \infty.
    \end{equation}
\end{itemize}
\end{theorem}
At last we want to clarify some terminology. The discussion above is focused on bounded linear functionals. These are related to states via the following definition: 

\begin{definition}[States on $C^\ast$-Algebras] Let $\cA$ be a $C^\ast$-algebra and $\phi$ a bounded linear functional on $\cA$. It is called a state if it is positive, i.e., $\phi(a^\dagger a) \geq 0$ for all $a \in \cA$ and normalized, i.e., $\norm{\phi} = 1$.
\end{definition}
If the algebra is represented on a Hilbert space, i.e., $\cA \subset \cB(\cH)$, then a density matrix $\r$ on this Hilbert space defines such an algebraic state via the vector functional which maps operators to their expectation value:
\begin{equation}
    a \mapsto \Tr(\r a).
\end{equation}
This includes pure states via their canonical density matrix $\r = \lvert\psi\rangle\langle\psi\rvert$. Throughout this work we will
usually not specify explicitly whether a state is a vector state, a mixed state in a concrete Hilbert space or an abstract state and refer to all of them simply as states. The concrete meaning of each instance of the word can be inferred from the context. In particular, by definition stabilizer states always refer to concrete vectors in finite-dimensional Hilbert spaces.

\subsection{Type classification}\label{app:type-recap}
Let $M\subset\cB(\cH)$ be a factor von Neumann algebra. This section
constitutes a summary of its type classification \cite{Blackadar}, see
\cite{Sorce:2023fdx} for a recent introduction aimed at physicists.

The type classification is based on the realization that different algebras
contain projections associated to subspaces of different sizes, and that these
sizes are basis-independent. For example, if $M=\cB(\cH)$, then
the projection $p_\y=\ket{\y}\bra{\y}$ is an element of $M$, but so is every
other one-dimensional projection. In particular, because $\cB(\cH)$ contains
all maps, we can consider the partial isometry
\begin{equation}
    v_{\p,\y}=\ket{\p}\bra{\y},
    \qquad
    \norm{\p}=1,
    \qquad
    \norm{\y}=1.
\end{equation}
Therefore, one has
\begin{equation}
    v_{\p,\y}^\dagger v_{\p,\y}=p_\y,
    \qquad
    v_{\p,\y}v_{\p,\y}^\dagger=p_\p.
\end{equation}
Thus, these two projections are equivalent relative to the algebra because we
can translate between them through the existence of a partial isometry. This
gets formalized in the following definitions.

\begin{definition}[Projection and order]
\thmref{cf.~II.3.3.1 in Ref.~\cite{Blackadar}}
A projection $p\in M$ is an operator with
\begin{equation}
    p=p^\dagger=p^2.
\end{equation}
For two projections $p,q\in M$, we say that $p\leq q$ if and only if $q-p\geq
0$, i.e.,
\begin{equation}
    \bra{\p}(q-p)\ket{\p}\geq 0
    \qquad\text{for all }\ket{\p}\in\cH.
\end{equation}
This also implies that $pq=qp=p$.
\end{definition}

\begin{definition}[Partial isometry]
\thmref{cf.~II.1.5.7 in Ref.~\cite{Blackadar}}
A partial isometry is an element $v\in M$ for which $v^\dagger v$ is a
projection. Equivalently, it is a unitary between its support and its range,
and the zero map on its kernel.
\end{definition}

\begin{definition}[Subequivalent projections]
\thmref{cf.~II.3.3.4 in Ref.~\cite{Blackadar}}
Let $p,q\in M$ be two projections. We say that $p$ is subequivalent to $q$,
written $p\precsim q$, if and only if there exists a partial isometry $v\in M$
such that
\begin{equation}
    v^\dagger v=p,
    \qquad
    vv^\dagger\leq q.
\end{equation}
\end{definition}

\begin{definition}[Equivalent projections]
\thmref{cf.~II.3.3.4 and III.1.1.9 in Ref.~\cite{Blackadar}}
Let $p,q\in M$ be two projections. We say that $p$ and $q$ are equivalent,
written $p\sim q$, if and only if there exists a partial isometry $v\in M$ such
that
\begin{equation}
    v^\dagger v=p,
    \qquad
    vv^\dagger=q.
\end{equation}
Equivalently, $p\precsim q$ and $q\precsim p$.
\end{definition}

\begin{definition}[Minimal and abelian projections]
\thmref{cf.~III.1.3.1 and III.1.3.10 in Ref.~\cite{Blackadar}}
A nonzero projection $p\in M$ is called minimal if its only subprojections are
$0$ and $p$. It is called abelian if $pMp$ is an abelian algebra.
\end{definition}

As can be understood for the case of finite-dimensional projections above, two
finite-dimensional projections are equivalent if they have the same
dimensionality. In particular, a $k$-dimensional projection cannot be
equivalent to a $k'$-dimensional projection with $k'<k$, because no partial
isometry that implements this equivalence can exist for unequal dimensions.
This is formalized for the general case via the following definition.

\begin{definition}[Finite and infinite projections]
\thmref{cf.~III.1.3.1 in Ref.~\cite{Blackadar}}
A projection $p\in M$ is finite if
\begin{equation}
    q\leq p,
    \qquad
    q\sim p
    \quad\to\quad
    q=p.
\end{equation}
    A projection $p\in M$ is called infinite if it is not finite, equivalently, if it is equivalent to
a proper subprojection of itself.
\end{definition}

\begin{definition}[Semifinite projection]
\thmref{cf.~III.1.4.2 in Ref.~\cite{Blackadar}}
A projection $p\in M$ is called semifinite if every nonzero subprojection of
$p$ contains a nonzero finite subprojection.
\end{definition}

\begin{definition}[Purely infinite projection]
\thmref{cf.~III.1.4.2 in Ref.~\cite{Blackadar}}
A projection $p\in M$ is called purely infinite if $p\neq0$ and it contains no
nonzero finite subprojections.
\end{definition}

\begin{definition}[Properly infinite projection]
\thmref{cf.~III.1.3.1 and III.1.3.5 in Ref.~\cite{Blackadar}}
A projection $p\in M$ is called properly infinite if $p\neq0$ and there are
orthogonal projections $p_1,p_2\leq p$ such that
\begin{equation}
    p_1\sim p,
    \qquad
    p_2\sim p.
\end{equation}
Equivalently, for every central projection $z\in\cZ(M)$, the projection $zp$
is either zero or infinite.
\end{definition}

These are the different types of projections that can appear within a von
Neumann algebra. We see that these are intrinsic features of it. In particular,
these turn out to be the defining characteristics of the types.

\begin{definition}[Types of factors]
\thmref{cf.~III.1.4.5--III.1.4.7 in Ref.~\cite{Blackadar}}
Let $M\subset\cB(\cH)$ be a factor von Neumann algebra.
\begin{itemize}
    \item[(i)] $M$ is called type I if every nonzero projection has a nonzero
    abelian subprojection. Equivalently, $M$ contains a minimal projection.
    \item[(ii)] $M$ is called type II if its identity projection is semifinite
    and it has no minimal projections. In particular, it is called type II$_1$
    if the identity is finite, and type II$_\infty$ otherwise.
    \item[(iii)] $M$ is called type III if it has no nonzero finite projections.
\end{itemize}
\end{definition}

\subsection{The GNS construction}\label{app:GNS}
The algebraic approach to quantum mechanics assumes the existence of an algebra
as the description of a system, or a collection of algebras in the form of a net. Thus the
primary objects are observables, rather than an underlying Hilbert space. The
Gelfand--Naimark--Segal (GNS) construction is a procedure which, given a
$C^\ast$-algebra $\cA$ and a state $\w: \cA \rightarrow \mathbb{C}$,
constructs a Hilbert space from this data. A formal discussion of the construction can be found in 
\cite[II.6.4]{Blackadar}. This procedure is the backbone of
much of the discussion in the main text and allows for the recovery of the more
conventional description of quantum mechanics where the Hilbert space is front
and center.
The GNS construction is used implicitly when there is a vacuum
state $\w$ and one has operators, which create "excitations on top of the
state" $\w$. For example, for the harmonic oscillator, we can consider the
algebra formed by the creation and annihilation operators $a, a^\dagger$ which
satisfy the standard canonical commutation relation \footnote{It is well known
    that $a,a^\dagger$ are unbounded operators, so their generated algebra does
    not have a well-defined norm. One remedies this by using exponentials of $a
    + a^\dagger, a- a^\dagger$ as the generators of an algebra but we will here
ignore this subtlety.}
\begin{equation}\label{eq:CCR}
    [a,a^\dagger] = 1.
\end{equation}
Then the Hilbert space of the harmonic oscillator is obtained by starting with
the vacuum vector $\ket{\W}$, which satisfies 
\begin{equation}
    a \ket{\W} = 0
\end{equation}
and has all other values defined via the CCR algebra~\eqref{eq:CCR}. In
particular, a basis of the Hilbert space is obtained essentially by taking the
identity $\id$ and $a^\dagger$ and mapping
\begin{equation}
    \id \mapsto \ket{\W},\  a^\dagger \mapsto a^\dagger \ket{\W} := \ket{1},\
    \hdots.
\end{equation}
From standard quantum mechanics we know that these vectors form a basis of the Hilbert
space of the harmonic oscillator.
We therefore see that the Hilbert space is obtained from the vacuum $\ket{\W}$ and the creation of "excitations on
top of it" by acting with elements of the algebra on this vacuum. 
The last ingredient in the GNS construction
is the so-called \emph{Gelfand ideal}. In the case of the harmonic
oscillator, the operator $a$ annihilates particles. Thus, if we take an
operator $O$ and consider $O + a$, acting with this on the vacuum will create
the same vector as just $O$, since the annihilation operator annihilates the vacuum and thus adding operators of this kind will not change the resulting
vector. More formally one has 
\begin{definition}
    Let $\cA$ be a $C^\ast\text{-algebra}$ and $\w$ a state. The set of
    operators $a \in \cA$ such that 
    \begin{equation}
        \w(a^\dagger a) = 0
    \end{equation}
    is the Gelfand-ideal $\cN_\w$.
\end{definition}
It can be easily shown that states satisfy a generalized Cauchy-Schwarz
inequality
\begin{equation}
    |\w(a^\dagger b)|^2 \leq \w(a^\dagger a) \w(b^\dagger b),
\end{equation}
from which we can infer that for every $a \in \cA$ and every $n \in \cN_\w$ one
has 
\begin{equation}
    0 \leq \w((an)^\dagger an) = \w(n^\dagger a^\dagger a n) \leq \norm{a}^2
    \w(n^\dagger n) = 0,
\end{equation}
where we used 
\begin{equation}
    a^\dagger a \leq \norm{a}^2 \id
\end{equation}
in the second equality. Therefore $\cN_\w$ is a left-ideal, i.e., if $n \in
\cN_\w$, then $an \in \cN_\w,\ \forall a \in \cA$.
In addition one has that $\cN_\w$ is a radical, i.e., for $a \in \cA$ and $n \in
\cN_\w$ one has 
\begin{equation}
    0 \leq |\w(n^\dagger a)|^2 \leq \w(n^\dagger n) \w(a^\dagger a) = 0,
\end{equation}
which implies that 
\begin{equation}\label{eq:radical}
    \w(n^\dagger a) = \w(a^\dagger n) = 0.
\end{equation}

The GNS construction now proceeds by considering the equivalence relation 
\begin{equation}
    a \sim b \Leftrightarrow a - b \in \cN_\w, 
\end{equation}
which defines equivalence classes $[a]$ for all $a \in \cA$. This set is
denoted by $\cA/\cN_\w$. We now assign a
vector to each equivalence class
\begin{equation}
    [a] \mapsto \ket{[a]},
\end{equation}
on which we define the inner product
\begin{equation}
    \braket{[b]}{[a]} = \w(b^\dagger a).
\end{equation}
Note that $\cN_\w$ is mapped to the zero vector, as it is orthogonal to every
vector in $\cA / \cN_\w$.
This inner product is well defined, since for two different representatives
$a,a',b,b'$ of $[a]$ and $[b]$ one has that $a - a' \in \cN_\w, b-b' \in
\cN_\w$ and thus
\begin{equation}
    \w((b')^\dagger a') - \w(b^\dagger a) = 0
\end{equation}
since by linearity this difference is a sum of terms of the form
\eqref{eq:radical}, which vanish. 
We can see that this inner product is
positive definite, since 
\begin{equation}
    \braket{[a]}{[a]} = 0
\end{equation}
implies $\w(a^\dagger a) = 0$ so that $a \in \cN_\w$, which in turn implies $[a] = 0$.
Thus we have obtained a vector space $\cV_\w \sim \cA/\cN_\w$ with a
positive definite inner product. Taking the norm completion in the norm induced
by the inner product then defines the GNS Hilbert space $\cH_\w$. 
This Hilbert space carries a natural representation $\pi_\w: \cA \rightarrow
\cB(\cH_\w)$ of $\cA$ defined by
\begin{equation}
    \pi_\w(a) \ket{[b]} = \ket{[ab]}.
\end{equation}
One can check again that, since $\cN_\w$ is a left-ideal, this representation
is well-defined. The vector associated to the equivalence class of the identity
$[\id]$ has a special role in that the linear functional arising from this vector recovers the state $\w$ since 
\begin{equation}
    \mel{[\id]}{\pi_\w(a)}{[\id]} = \braket{[\id]}{[a]} = \w(\id^\dagger a) =
    \w(a).
\end{equation}
Thus, the "vacuum" $\ket{[\id]}$ represents the state $\w$ in $\cH_\w$ and all
vectors in $\cH_\w$ are obtained by acting on $\ket{[\id]}$ with operators in
$\cA$ since 
\begin{equation}
    \ket{[a]} = \ket{[a \id]} = \pi_\w(a) \ket{[\id]}.
\end{equation}
In this precise sense $\cH_\w$ contains vectors that are "excitations" on top
of the "vacuum" induced by the state $\w$.
The physical importance of the GNS construction is that each vector in $\cH_\w$ models a
different physical situation in which the degrees of freedom that are described
by the algebra $\cA$ can be found in. In particular, the GNS Hilbert space
$\cH_\w$ is intimately related to the entanglement structure of the state. 

\subsection{Representations of algebras}
    Throughout this work we will deal with algebras on different levels. We will have algebras abstractly as $C^\ast$ algebras of operators. Then we will represent them on specific Hilbert spaces, complete them etc. This naturally gives rise to the question, when two algebras that arise from different representations are equivalent. Here we will collect some basic theory concerning these questions.

\begin{definition}[Dual Banach space]
Let $B$ be a Banach space. Then $B^*$ is the set of bounded linear functionals $\ell$
on $B$ with the norm
\begin{equation}
    \norm{\ell}_{B^*}
    =
    \sup_{\norm{b}_B \leq 1} \abs{\ell(b)}.
\end{equation}
\end{definition}

\begin{definition}[Invariant subspace associated with a representation]
Let $\cA$ be a $C^\ast$-algebra and
$\pi:\cA \rightarrow \cB(\cH)$ a non-degenerate representation of $\cA$. Set
$M_\pi = \pi(\cA)''$ and define
\begin{equation}
    \cF_\pi(\cA)
    :=
	    \{\P \circ \pi : \P \in (M_\pi)_*\}
	    \subset \cA^*.
\end{equation}
This is called the invariant subspace of $\cA^*$ associated with $\pi$.
\end{definition}

\begin{lemma}[Closedness of $\cF_\pi(\cA)$]\label{lem:takesaki-folium-closed}
\thmref{cf.~Ch.~III, below Def.~2.10, p.~125 in Ref.~\cite{TakesakiI}}
Let $\cA$ be a $C^\ast$-algebra and
$\pi:\cA \rightarrow \cB(\cH)$ a representation of $\cA$. Then
$\cF_\pi(\cA)$ is a closed subspace of $\cA^*$.
\end{lemma}
To later compare the subspaces associated to different representations we need the following lemma.
\begin{lemma}[Vector-functional characterization of $\cF_\pi(\cA)$]\label{lem:linear-folium}
Let $\cA$ be a $C^\ast$-algebra and
$\pi:\cA \rightarrow \cB(\cH)$ a non-degenerate representation of $\cA$.
Then $\cF_\pi(\cA)$ is the closure of the linear span of
$\omega_{\xi,\eta}^\pi$ ~for $\xi,\eta \in \cH$, in the norm of the dual space:
\begin{equation}
	    \cF_\pi(\cA)
	    =
    \overline{\operatorname{span}}
    \{\omega_{\xi,\eta}^\pi : \xi,\eta \in \cH\}^{\norm{\cdot}_{\cA^*}},
\end{equation}
where
\begin{equation}
	    \omega_{\xi,\eta}^\pi(a) = \mel{\xi}{\pi(a)}{\eta},
	    \quad a \in \cA.
\end{equation}
\end{lemma}

\begin{proof}
Let $\P \in (M_\pi)_*$. By
Theorem~\ref{thm:blackadar-normal-vector}, $\P$ can be written as
\begin{equation}
    \P(x) = \sum_{n=1}^{\infty} \mel{\psi_n}{x}{\eta_n},
    \quad x\in M_\pi,
\end{equation}
with
\begin{equation}
    \sum_{n=1}^{\infty}\norm{\psi_n}^2 < \infty,
    \quad
    \sum_{n=1}^{\infty}\norm{\eta_n}^2 < \infty.\label{eq:square-summable}
\end{equation}
We want to show that $\P \circ \pi$ can be written as the norm limit of finite
sums of matrix elements. We now show that the finite sums converge in norm to the full functional. For all
$a\in\cA$ with $\norm{a}\leq 1$, one has
\begin{align}
    \abs{
        (\P\circ\pi)(a)
        -
        \sum_{n=1}^{N}\omega_{\psi_n,\eta_n}^{\pi}(a)
    } \notag
    &=
    \abs{
        \sum_{n>N}\mel{\psi_n}{\pi(a)}{\eta_n}
    } \notag\\
    & \leq
    \sum_{n>N}\norm{\psi_n}\norm{\pi(a)\eta_n} \notag\\
    & \leq
    \left(\sum_{n>N}\norm{\psi_n}^2\right)^{1/2}
    \left(\sum_{n>N}\norm{\pi(a)\eta_n}^2\right)^{1/2} \notag\\
    & \leq
    \norm{a}
    \left(\sum_{n>N}\norm{\psi_n}^2\right)^{1/2}
    \left(\sum_{n>N}\norm{\eta_n}^2\right)^{1/2} \notag\\
    & \leq
    \left(\sum_{n>N}\norm{\psi_n}^2\right)^{1/2}
    \left(\sum_{n>N}\norm{\eta_n}^2\right)^{1/2}.
\end{align}
The second line used the triangle inequality and Cauchy-Schwarz inequality in $\cH$, the third line
used the Cauchy-Schwarz inequality in the Hilbert space $\ell^2$ of
square-summable sequences. The fourth line used the definition of the operator
norm and that representations of $C^\ast$-algebras are contractive
\cite[II.6.1.2, p.~106]{Blackadar}. In the last inequality we used
$\norm{a}\leq 1$. This is a uniform bound for all
$a$ in the $C^\ast$-algebra with $\norm{a}\leq 1$. The right-hand side vanishes
as $N\rightarrow\infty$, since the series in~\eqref{eq:square-summable} converge. Therefore the linear functional defined by the partial
sums converges in norm to the full linear functional. We 
therefore know that
$\cF_\pi(\cA) \subset \overline{\operatorname{span}}
    \{\omega_{\xi,\eta}^\pi : \xi,\eta \in \cH\}^{\norm{\cdot}_{\cA^*}}$.
Conversely, each linear functional of the form $\omega_{\psi,\eta}^{\pi}$ is
the restriction along $\pi$ of a normal functional on $M_\pi$, because this is
the case of the definition of normal functionals for a sum with just a single
term. Therefore, finite sums of linear functionals are
also in $\cF_\pi(\cA)$. So $\cF_\pi(\cA) \supset \operatorname{span}
\{\omega_{\xi,\eta}^\pi : \xi,\eta \in \cH\}$.
Lemma~\ref{lem:takesaki-folium-closed} asserts that $\cF_\pi(\cA)$ is a closed
subspace of $\cA^*$, so
\begin{equation}
    \overline{\operatorname{span}}
    \{\omega_{\xi,\eta}^\pi : \xi,\eta \in \cH\}^{\norm{\cdot}_{\cA^*}}
    \subset
    \cF_\pi(\cA)
    \subset
    \overline{\operatorname{span}}
    \{\omega_{\xi,\eta}^\pi : \xi,\eta \in \cH\}^{\norm{\cdot}_{\cA^*}},
\end{equation}
which implies
\begin{equation}
    \cF_\pi(\cA)
    =
    \overline{\operatorname{span}}
    \{\omega_{\xi,\eta}^\pi : \xi,\eta \in \cH\}^{\norm{\cdot}_{\cA^*}},
\end{equation}
\end{proof}

\begin{definition}[Quasi-equivalent representations]
\thmref{cf.~Ch.~III, Def.~2.10 in Ref.~\cite{TakesakiI}}
Let $\pi_1:\cA \rightarrow \cB(\cH_1)$ and
$\pi_2:\cA \rightarrow \cB(\cH_2)$ be two non-degenerate representations of a $C^\ast$-algebra
$\cA$. The representations $\pi_1$ and $\pi_2$ are called quasi-equivalent if
and only if there exists a $\sigma$-weakly bicontinuous $\ast$-isomorphism between their generated von Neumann algebras $M_{\pi_i}$
\begin{equation}
    \rho:M_{\pi_1} \rightarrow M_{\pi_2}
\end{equation}
such that
\begin{equation}
    \rho(\pi_1(a)) = \pi_2(a),
    \quad a\in\cA.
\end{equation}
\end{definition}
\begin{proposition}[Quasi-equivalence from closed invariant subspaces]\label{prop:equivalence-representations}
\thmref{cf.~Ch.~III, Prop.~2.12, p.~126 in Ref.~\cite{TakesakiI}}
Let $\pi_1:\cA \rightarrow \cB(\cH_1)$ and
$\pi_2:\cA \rightarrow \cB(\cH_2)$ be two representations of a $C^\ast$-algebra
$\cA$. The two representations are quasi-equivalent, equivalently the two von
Neumann algebras generated by $\cA$ relative to $\pi_1$ and $\pi_2$ are
equivalent over $\cA$, if and only if
\begin{equation}
    \cF_{\pi_1}(\cA) = \cF_{\pi_2}(\cA).
\end{equation}
\end{proposition}
The last proposition gives us a checkable criterion for whether the von Neumann
algebras generated by $\cA$ for two representations are isomorphic.

\section{Infinite stabilizer systems}\label{app:lattice_systems}
This appendix introduces our main framework of stabilizer systems on lattices. The main result is a proof that a given stabilizer system on an infinite lattice defines a unique state that describes the projection onto the local code spaces.
Let $\cL$ be a countable lattice with vertices $V$ and for each vertex a qubit
Hilbert space $\cH_v = \mathbb{C}^2$. For each finite subset $\L \Subset \cL$
we define the associated Hilbert space
\begin{equation}
   \cH_\L = \bigotimes_{v \in \L} \cH_v,
\end{equation}
and the associated \emph{local algebra of observables}
\begin{equation}
    \cA_\L = \cB(\cH_\L) \sim M_{2^{|\L|}(\bbC)}.
\end{equation}
If $\Lambda \subset \Delta$ , we identify $\cA_\L$ with the subalgebra
$\cA_\L \otimes \id_{\D \setminus \L} \subset \cA_\D$, i.e., we define
the equivalence relation 
\begin{equation}\label{eq:lattice_embedding}
    \cA_\L \sim \cA_\L \otimes \id_{\D \setminus \L}
\end{equation}
We denote the complement of $\L$ by $\L^{c} = \cL \setminus \L$.
We say that an operator $O$ is a \emph{local operator}, if
$O\in\cA_F$ for some finite $F\Subset\cL$, and write
$\operatorname{supp}(O)\subseteq F$.
We define
\begin{equation}
    \cA_{\text{loc}}:=\bigcup_{\L\Subset\cL}\cA_{\L},
\end{equation}
as the algebra of local observables. Define the \emph{quasi-local algebra of
observables} as
\begin{equation}
    \cA:=\overline{\cA_{\text{loc}}}^{\|\cdot\|},
\end{equation}
the norm completion of the union of local algebras, where algebras are
identified according to Eq.~\eqref{eq:lattice_embedding}. We will in the
following always implicitly identify operators according to Eq.~\eqref{eq:lattice_embedding}, when we talk about unions or take elements that
live in one set and perform set-theoretic operations on them in larger sets.
More formally, we would have to define an $\ast$-homomorphism $\a_{\L}:
\cA_{\L}\rightarrow \cA$ explicitly that maps each operator to its equivalence
class under~\eqref{eq:lattice_embedding}. We will follow a rather
informal presentation, where we will use the symbol $\cA_F$ for $F \Subset \cL$ a
finite set, both as the finite-dimensional matrix algebra, as well as the
subalgebra in $\cA$. Which one is meant should be clear from context. Since in
a more formal treatment, we would have an $\ast$-homomorphism from $\cA_{\L}$
to its image in $\cA$, all inequalities that we prove for the finite
dimensional case below, also hold for the image. This follows since
$\ast$-homomorphisms are positive maps \cite[Section 4.2]{KadisonRingrose1}, i.e., since $P \geq 0 \Rightarrow  \a(P) \geq 0$, we have that if 
\begin{equation}
    P' \geq P \Leftrightarrow P'-P \geq 0
\end{equation}
then $\a(P'-P) = \a(P') -\a(P) \geq 0$ which implies the preservation of the
inequality in the image.

For a finite region $\L \Subset \cL$, we define the \emph{local
Pauli operators} $\cP_\L \subset \cA_\L$ to be given by operators of the form 
\begin{equation}
    O = \bigotimes_{i \in \L} p_i, \quad p_i \in \{X,Y,Z,\id\}.
\end{equation}
The \emph{local Pauli group} $\cG^{\cP}_\L$ is the group generated by elements
of $\cP_\L$, i.e., all operators obtained by taking finite products of
elements of $\cP_\L$. The global Pauli group $\cG^{\cP}$ is the group 
\begin{equation}
    \cG^{\cP} = \bigcup_{\L\Subset\cL} \cG^{\cP}_\L.
\end{equation}
These notions allow us to define 
\begin{definition}[Consistent stabilizer system]
A \emph{stabilizer system} is a set $\cS$, whose elements are mutually
commuting Pauli operators with finite support. A \emph{consistent stabilizer
system} is a stabilizer system where no product of its elements equals
$-\id$.     
\end{definition}
In the following, we assume that all stabilizer systems have finite
support, i.e., every stabilizer is supported on a finite set of sites.

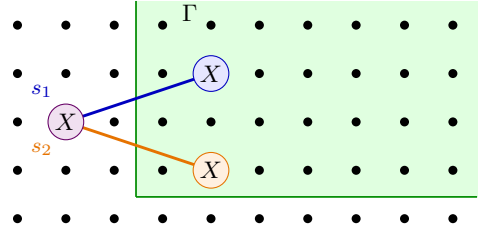
\begin{figure}
    \begin{tikzpicture}[x=.64cm,y=.64cm,
    site/.style={circle,fill=black,inner sep=1.2pt},
    support/.style={line width=1.1pt}]
  \fill[green!12] (2.45,.45) rectangle (9.5,4.5);
  \draw[green!55!black,line width=.65pt] (2.45,.45) -- (9.5,.45);
  \draw[green!55!black,line width=.65pt] (2.45,.45) -- (2.45,4.5);
  \foreach \x in {0,...,9}
    \foreach \y in {0,...,4}
      \node[site] at (\x,\y) {};
  % Both stabilizers act by X at their common exterior site, but their
  % interior supports are disjoint.
  \draw[support,blue!75!black] (1,2) -- (4,3);
  \draw[support,orange!90!black] (1,2) -- (4,1);
  \node[circle,draw=violet!80!black,fill=violet!12,inner sep=1.1pt] at (1,2) {$X$};
  \node[circle,draw=blue!75!black,fill=blue!10,inner sep=1.1pt] at (4,3) {$X$};
  \node[circle,draw=orange!90!black,fill=orange!12,inner sep=1.1pt] at (4,1) {$X$};
  \node[blue!75!black,anchor=east,font=\footnotesize] at (.9,2.65) {$s_1$};
  \node[orange!90!black,anchor=east,font=\footnotesize] at (.9,1.45) {$s_2$};
   \node[black,anchor=east] at (3.9,4.25) {$\G$};
\end{tikzpicture}%
    \caption{An example of two $X X$-type stabilizers $s_1,s_2$, both of which are not supported in $\G$ but whose product is localized in $\G$ and thus contributes to the local stabilizer group of the region $\G$.}
    \label{fig:nontrivial-local-stabilizer-group-element}
\end{figure}

The \emph{stabilizer group} $\cG^{\cS}$ is the group generated by finite
products of elements of $\cS$. For a finite region $F\Subset\cL$, the
\emph{local stabilizer group} is
\begin{equation}\label{eq:local_stabilizer_group}
    \cG^{\cS}_{F}=\cG^{\cS}\cap\cA_F.
\end{equation}
Note that this contains all elements of $\cG^\cS$ that are localized in $\cA_F$, not just elements that arise from the product of stabilizers in $F$. An example is illustrated in Figure \ref{fig:nontrivial-local-stabilizer-group-element}.

For a finite region $F\Subset\cL$, we can define the \emph{projection onto
the local code subspace} as
\begin{equation}\label{eq:projection_local}
    P_F=\frac{1}{|\cG^{\cS}_{F}|}\sum_{g\in\cG^{\cS}_{F}}g.
\end{equation}
This projection satisfies
\begin{equation}
    gP_F=P_F,
    \qquad g\in\cG^{\cS}_{F},
\end{equation}
and therefore projects onto the common $+1$ eigenspace of all stabilizers.
This defines a local density matrix
\begin{equation}
    \rho_F=\frac{1}{\operatorname{Tr}(P_F)}P_F.
\end{equation}
We can compute
\begin{equation}
    \operatorname{Tr}(P_F)=\frac{2^{|F|}}{|\cG^{\cS}_{F}|},
\end{equation}
because all summands that are not the identity are traceless: products of
Pauli operators are proportional to Pauli operators, and all non-identity Pauli
operators are traceless. Therefore, only the identity contributes, which gives the factor
$2^{|F|}$. We therefore find that
\begin{equation}
    \rho_F=\frac{1}{2^{|F|}}\sum_{g\in\cG^{\cS}_{F}}g.
    \label{eq:local_density_full}
\end{equation}

We now note that if $F\subset F'$ are finite subsets of $\cL$, we have
\begin{equation}\label{eq:rho-compatibility}
    \rho_F=\operatorname{Tr}_{F'\setminus F}(\rho_{F'}).
\end{equation}
This can be seen by the fact that
\begin{equation}
    2^{-|F'|}2^{|F'\setminus F|}=2^{-|F|},
\end{equation}
and that
\begin{equation}
    \operatorname{Tr}_{F'\setminus F}(g)\neq0
\end{equation}
if and only if $g$ is of the form
\begin{equation}
    g=g'\otimes\id_{F'\setminus F},
    \qquad g'\in\cG^{\cS}_{F}.
\end{equation}
This in turn implies
\begin{align}
    \operatorname{Tr}_{F'\setminus F}(\rho_{F'})
    &=\frac{1}{2^{|F'|}}
      \sum_{g\in \cG^{\cS}_{F'}}
      \operatorname{Tr}_{F'\setminus F}(g) 
      \\
      \nonumber
    &=\frac{1}{2^{|F'|}}2^{|F'\setminus F|}
      \sum_{g\in\cG^{\cS}_{F}}g \\
       \nonumber
    &=\frac{1}{2^{|F|}}\sum_{g\in\cG^{\cS}_{F}}g
      =\rho_F.
       \nonumber
\end{align}
Here we used that for $g\in\cG^{\cS}_{F}$,
\begin{equation}
    g\otimes\id_{F'\setminus F}\in \cG^{\cS}_{F'}.
\end{equation}
As follows from the discussion above, all elements not of that form vanish
under the trace over $F'\setminus F$.

This allows us to define a state that implements this projection on the
quasi-local algebra via
\begin{proposition}[Consistent stabilizer systems define a unique state]\label{prop:define_state}
    Let $\cS$ be a consistent stabilizer system on a lattice $\cL$. Then the
    state defined by the reduced density matrices $\rho_F$ above defines a
    unique state on $\cA$.
\end{proposition}
\begin{proof}
    The local density matrices already define the state
    \begin{equation}
        \w(a):=\operatorname{Tr}_F(\rho_Fa),
        \qquad a\in\cA_F.
    \end{equation}
    The consistency relation
    \begin{equation}
        \rho_F=\operatorname{Tr}_{F'\setminus F}(\rho_{F'})
    \end{equation}
    proves that the state is independent of the representation. Thus, if
    $a$ is considered as an element of $\cA_F$ and of $\cA_{F'}$, then
    \begin{align}
        \operatorname{Tr}_{F'}\bigl(\rho_{F'}(a\otimes\id_{F'\setminus F})\bigr)
        &=\operatorname{Tr}_F\bigl(\operatorname{Tr}_{F'\setminus F}(\rho_{F'})a\bigr) \\
        &=\operatorname{Tr}_F(\rho_Fa).
        \nonumber
    \end{align}
    Therefore, the state is well-defined on elements in $\cA_{\text{loc}}$.
    Since by definition $\cA_{\text{loc}}$ is a norm-dense subset of $\cA$,
    the functional $\w$ extends uniquely to a functional on $\cA$ with the same
    norm by the B.L.T. Theorem \cite[Thm.~I.7]{ReedSimon1972}.

    To show that it is a state, we also need to show positivity and
    normalization. We have $\w(\id) = 1$, since this holds for the identity on
    each finite subset. To prove positivity assume that $a\in\cA$ is positive. By
    \cite[Proposition~II.3.1.2(vii)]{Blackadar}, $a$ has a unique positive
    square root $c$, i.e.,
    \begin{equation}
        a=c^\dagger c.
    \end{equation}
    Approximate $c$ by $c_n\in\cA_{F_n}\subset\cA_{\text{loc}}$. Since multiplication is norm continuous we have that $c^\dagger_n c_n$ converges to $c^\dagger c$ in norm.
    Since $\w$ is a bounded and thus a continuous linear functional, we have
    \begin{equation}
        \w(a)=\lim_{n\to\infty}\w(c_n^*c_n).
    \end{equation}
    Furthermore,
    \begin{equation}
        \w(c_n^*c_n)
        =\operatorname{Tr}_{F_n}(\rho_{F_n}c_n^*c_n)\geq0,
    \end{equation}
    which implies $\w(a)\geq0$. 
\end{proof}
\section{No type III factors in stabilizer states}\label{app:notypeIII}
Now that we have a state, we can consider the Hilbert space $\cH_\w$ which
is the GNS Hilbert space for $\cA$ with respect to the state $\w$
\cite[II.6.4]{Blackadar}, which provides a representation $\pi_\w : \cA
\rightarrow \cB(\cH_\w)$ and a vector $\ket{\W} \in \cH_\w$ such that 
\begin{equation}
    \w(a) = \bra{\W}\pi_\w(a) \ket{\W},\quad \forall a \in \cA.
\end{equation}
For a subregion $\G \subset \cL$, define the algebra of finite local
operators in the GNS representation by
\begin{equation}
    \cU_F:=\pi_\w(\cA_F),
    \qquad
    \cU_{\G,\mathrm{fin}} = \bigcup_{F \Subset \G} \cU_F.
\end{equation}
This is a $\ast$-algebra: if two elements are supported in finite regions $F$
and $E$, then their sum, product, and adjoint belong to
$\cU_{F\cup E}$. We define the local $C^\ast$-algebra in the GNS Hilbert
space by
\begin{equation}
    \cU_{\G} = \overline{\cU_{\G,\mathrm{fin}}}^{\norm{.}}.
\end{equation}
We then consider the von Neumann algebra
\begin{equation}
    M_\G = \cU_\G''.
\end{equation}
This is our stand-in for the algebra of local operators that should describe
all physics that is accessible in the state $\w$.
Our goal is to study the central decomposition of $M_\G$ and show that no type
III factors appear there. We will do so by following the proof strategy of
\cite{Ogata:2022qdm}, by taking an exhaustion of $\G$ from
finite sets
\begin{equation}
    \G_1 \subset \G_2 \subset \hdots, \quad \G = \bigcup_{n} \G_n.
\end{equation}
and consider the local projections $P_{\G_n}$ from~\eqref{eq:projection_local}. We show that these converge to a
projection $q$ where $q M_\G q$ is finite and that this finiteness is imprinted on the central decomposition of $M_\G$. This implies that each factor in the
central decomposition contains a finite projection, which is incompatible with
it being type III. We will demonstrate finiteness directly by using the state
$\w$, which will form a tracial state on $q M_\G q$. Physically, the traciality
arises, because the local state is proportional to a projection onto the local
code space and $q$ locally reduces to this projection, so that $q M_\G q$ contains operators 
that map the code space to itself, and the resulting tracial state is just the standard
trace over the code space.
We first need to show that the sequence $\G_n$ defines a decreasing set of
projections, i.e.,
\begin{lemma}[Monotonicity of local support projections]\label{lema:proj_inequality}
    Let $\{\G_n \}_{n \in \mathbb{N}}$ be an exhaustion of $\G$ by finite sets
    and let $P_n:=P_{\G_n}$ be the local projection of
    Eq.~\eqref{eq:projection_local}, write
    $p_n:=\pi_\w(P_n)$. Then we have
    \begin{equation}
    p_{n} \geq p_{n+1},
\end{equation}
\end{lemma}
For this proof we need the following lemma.
\begin{lemma}[Decrease of support projections under partial trace]
    Let $\cH_{1,2} = \cH_1 \otimes \cH_2$ be a tensor product of finite-dimensional
    Hilbert spaces, $\r_{1,2}$ a density matrix of $\cH_{1,2}$ with marginal
    $\r_1 = \Tr_{2} \r_{1,2}$. For a positive operator $\r$, let $s(\r)$ be the projection onto its support, i.e., if 
    \begin{equation}
        \r = \sum_{i = 1}^{\mathrm{rank \r}} \l_i \ket{i}\bra{i}
    \end{equation}
    then
    \begin{equation}
        s(\r) = \sum_{i = 1}^{\mathrm{rank \r}} \ket{i}\bra{i}.
    \end{equation}
    Then we have 
    \begin{equation}
        s(\r_1) \otimes \id_2 \geq s(\r_{1,2}).
    \end{equation}
\end{lemma}

\begin{proof}
    Let 
    \begin{equation}
        e = (\id_1 - s(\r_1)) \otimes \id_2
    \end{equation}
    Then
    \begin{equation}
        \Tr e \r_{1,2} = \Tr (\r_1 (\id_1 - s(\r_1))) = 0,
    \end{equation}
    by the definition of $s(\r_1)$. Since $e$ is a projection, thus $e^2 = e$, we have that
    \begin{equation}
        0 = \Tr e \r_{1,2} = \Tr \r^{1/2}_{1,2} e e \r^{1/2}_{1,2} = \norm{e
        \r^{1/2}_{1,2}}^{2}_2,
    \end{equation}
    is the square of the Hilbert-Schmidt norm, which implies that 
    \begin{equation}
        e\r^{1/2}_{1,2} = 0,
    \end{equation}
    so that the projection on the support of $\r^{1/2}_{1,2}$, and in turn also the projection on the support of
    $\r_{1,2}$, must be contained in $1-e =s(\r_1) \otimes \id_2$, which is
    exactly the statement to be proven,
    \begin{equation}
        s(\r_1) \otimes \id_2 \geq s(\r_{1,2}).
    \end{equation}
\end{proof}

\begin{proof}[Proof of Lemma \ref{lema:proj_inequality}]
    Since by Eq.~\eqref{eq:rho-compatibility} $\r_{\G_n} = \Tr_{\G_{n+1} \setminus \G_{n}} \r_{\G_{n+1}}$, the previous lemma implies
    \begin{equation}
        s(\r_{\G_n}) \otimes \id_{\G_{n+1} \setminus \G_{n}} \geq s(\r_{\G_{n+1}}).
    \end{equation}
    By the definition of $P_{\G_n}$ we have $s(\r_{\G_n})=P_{\G_n}$. Thus the canonical inclusions give
    $P_n \geq P_{n+1}$, and since
    the GNS representation is a $\ast$-homomorphism and therefore also preserves
    inequalities, the original statement follows.
\end{proof}

We thus have
\begin{lemma}[Limiting code-space projection]\label{lem:q_exists}
    The sequence $p_n$ has a strong limit $q \in M_\G$ and we have 
    \begin{equation}\label{eq:proj_stabilizer}
        q \ket{\W} = \ket{\W}.
    \end{equation}    
\end{lemma}
\begin{proof}
    By \cite[I.5.1.3]{Blackadar} a decreasing net of projections in $\cB(\cH)$ has
a strong limit, i.e.,
\begin{equation}
    p_n \xrightarrow{\mathrm{s}} q
\end{equation}
for some $q$. Since $M_\G$ is a von Neumann algebra, by the bicommutant theorem
\cite[I.9.1.2]{Blackadar} it is strongly closed. Since $p_n \in M_\G, \forall
n$ we have that its strong limit also belong to $M_\G$, so that $q \in M_\G$.
To prove $q \ket{\W} = \ket{\W}$ we consider the vector
\begin{equation}
    v = (1-q) \ket{\W}
\end{equation}
and show that $\norm{v} = 0$. Since $q$ is the strong limit, we know that 
\begin{equation}
    v_n = (1-p_n) \ket{\W}
\end{equation}
converges to $v$ in norm. But
\begin{equation}
    \norm{v_n}^2 = \bra{\W}(1-p_n)^2 \ket{\W} = \bra{\W}(1-p_n)\ket{\W} = \w(1)
    - \w(P_n).
\end{equation}
Furthermore,
\begin{equation}
    \w(P_n) = \Tr_{\G_n}(\r_{\G_n}P_n)
    = \Tr_{\G_n}(\r_{\G_n}) = 1,
\end{equation}
so that $\norm{v_n}^2 = 0$ and thus $v_n = 0$, which implies by
norm-convergence that $v = 0$, which implies Eq.~\eqref{eq:proj_stabilizer}.
\end{proof}
We now want to embark on the proof that for the algebra $qM_\G q$, the vector
$\ket{\W}$ induces a tracial state, i.e.,
\begin{theorem}[Faithful normal trace on the  corner]\label{thm:faithful_normal_trace}
    The functional
    \begin{equation}
        \tau_q(x) = \bra{\W} x \ket{\W},\quad x \in q M_\G q
    \end{equation}
    is a faithful normal tracial state on $q M_\G q$.
\end{theorem}
We will first prove that it is a normal tracial state by showing it to be tracial for increasingly large subalgebras of $qM_\G q$ in Lemmas \ref{lem:finite-trace} and \ref{lem:cstar_trace}, and then proving that it is a normal tracial state on the von Neumann algebra in Proposition \ref{prop:normal-trace}. We will prove that it is faithful in Proposition \ref{prop:faithful} which establishes Theorem \ref{thm:faithful_normal_trace}.

\begin{lemma}[Traciality on finite local corners]\label{lem:finite-trace}
    For every $a,b \in \cU_{\G,\mathrm{fin}}$,
\begin{equation}\label{eq:local_trace}
  \tau_q\bigl((qaq)(qbq)\bigr)
  =
  \tau_q\bigl((qbq)(qaq)\bigr).
\end{equation}
\end{lemma}

\begin{proof}
Choose $m$ large enough such that $a,b\in\cU_{\G_m}$, and choose
$\widetilde a,\widetilde b\in\cA_{\G_m}$ with
$a=\pi_\w(\widetilde a)$ and $b=\pi_\w(\widetilde b)$. For every $n\geq m$,
regard $\widetilde a$ and $\widetilde b$ as elements of $\cA_{\G_n}$. Then
\begin{align}\label{eq:local_state}
  \bra{\W}p_nap_nbp_n\ket{\W}
  &= \omega\bigl(P_n\widetilde aP_n\widetilde bP_n\bigr) \\
  \nonumber
  &= \frac{|\cG^{\cS}_{\G_n}|}{2^{|\G_n|}}
  \Tr_{\G_n}\bigl(P_n\widetilde aP_n\widetilde bP_n\bigr).
\end{align}
Since
\begin{equation}
  P_n\widetilde aP_n\widetilde bP_n
  =(P_n\widetilde aP_n)(P_n\widetilde bP_n),
\end{equation}
cyclicity of the finite-dimensional matrix trace gives
\begin{equation}
  \Tr_{\G_n}\bigl(P_n\widetilde aP_n\widetilde bP_n\bigr)
  =\Tr_{\G_n}\bigl(P_n\widetilde bP_n\widetilde aP_n\bigr).
  \label{eq:local_traciality}
\end{equation}

Since it is a set of projections, the sequence $(p_N)$ is uniformly bounded and
converges strongly to $q$. Since $a,b$ are bounded operators, the set of
operators $p_n a p_n, p_n b p_n$ is uniformly
bounded by $\max(\norm{a},\norm{b})< \infty$.
By \cite[I.3.2.1]{Blackadar} multiplication is jointly strongly continuous on bounded sets, so
\begin{equation}
  p_Nap_Nbp_N
  \xrightarrow{\mathrm{s}}
  qaqbq.
\end{equation}
The same holds with $a$ and $b$ interchanged.  Taking the limit
of~\eqref{eq:local_state} and using~\eqref{eq:local_traciality}
proves~\eqref{eq:local_trace}.
\end{proof}
Now that we have traciality on simple operators, we can study it on 

\begin{lemma}[Traciality on the norm closure]\label{lem:cstar_trace}
For every $a,b\in \cU_\G$,
\begin{equation}\label{eq:Cstar-trace}
  \tau_q\bigl((qaq)(qbq)\bigr)
  =
  \tau_q\bigl((qbq)(qaq)\bigr).
\end{equation}
\end{lemma}

\begin{proof}
Since $\cU_\G$ is the norm closure of $\cU_{\G,\mathrm{fin}}$, we can choose sequences $a_k,b_k\in\cU_{\G,\mathrm{fin}}$ such that
\begin{equation}
  a_k \xrightarrow{\norm{\cdot}} a,
  \qquad
  b_k \xrightarrow{\norm{\cdot}} b.
\end{equation}
Consider the operators
\begin{equation}
  X_k:=qa_kq,
  \quad
  Y_k:=qb_kq,
  \quad
  X:=qaq,
  \quad
  Y:=qbq.
\end{equation}
Since $\norm{q} \leq 1$ and the operator norm is submultiplicative as a
$C^\ast$-norm
\cite[II.1.1.1]{Blackadar} the map $x \mapsto q x q$ satisfies
\begin{equation}
    \norm{q x q} \leq \norm{q} \norm{x} \norm{q} \leq \norm{x},
\end{equation}
i.e., it is contractive. Therefore we have 
\begin{equation}
  X_k \xrightarrow{\norm{\cdot}} X,
  \qquad
  Y_k \xrightarrow{\norm{\cdot}} Y.
\end{equation}
By submultiplicativity of the norm, we also have that 
\begin{equation}
    X_kY_k \xrightarrow{\norm{\cdot}} XY,
    \qquad
    Y_kX_k \xrightarrow{\norm{\cdot}} YX.
\end{equation}
Since $\tau_q$ is a bounded linear functional, it is continuous with respect to the operator
norm. Therefore
\begin{equation}
  \tau_q(XY)
  =\lim_k\tau_q(X_kY_k)
  =\lim_k\tau_q(Y_kX_k)
  =\tau_q(YX).
\end{equation}
\end{proof}
Now that we know that the state is tracial on $q \cU_\G q$, we want to extend this result to the full von Neumann
algebra $M_\G$.
\begin{proposition}[Normal trace on the full corner]\label{prop:normal-trace}
The state $\tau_q$ defines a normal tracial state on $q M_\G q$.
\end{proposition}

\begin{proof}
By Lemma \ref{lem:q_exists} $q \in M_\G$ and \cite[I.9.1.4]{Blackadar} implies
that $qM_\G q$ is a von
Neumann algebra considered on the Hilbert space $q \cH_\w$.
To prove the assertion, we need to first prove that $q \cU_\G q$ is strongly
dense in $q M_\G q$, to approximate the functional on it.
For that, let $x \in M_\G$. By Kaplansky's density theorem
\cite[I.9.1.3]{Blackadar}, there exists a uniformly bounded net $(x_i)_{i\in I}$ in $\cU_\G$ with $x_i \xrightarrow{\sigma\text{-}\mathrm{s}} x$ and hence
strongly. By the same argument as in the previous lemma then
\begin{equation}
    q x_i q \xrightarrow{\sigma\text{-}\mathrm{s}} q x q,
\end{equation}
strongly which proves that $q \cU_\G q$ is strongly dense in $q M_\G q$. The
same argument implies that for every $a,b \in M_\G$ there exist uniformly bounded nets $(a_i)_{i\in I},(b_j)_{j\in J}$ in $\cU_\G$ such that
\begin{equation}
   q a_i q \xrightarrow{\mathrm{s}} q a q,
   \qquad
   q b_j q \xrightarrow{\mathrm{s}} q b q.
\end{equation}
To argue for the convergence of the product, we can introduce a joint index $k$\footnote{For example $k
    = (i,j)$ so that $a_{(i,j)} = a_i, b_{(i,j)} = b_j$ and $(i',j') \preceq (i,j)
    \Leftrightarrow i' \preceq i \land j' \preceq j$},
so that by the same argument as in the proof of Lemma \ref{lem:finite-trace}
\begin{equation}
    q a_i q b_j q \xrightarrow{\mathrm{s}} q a q b q,
    \qquad
    q b_j q a_i q \xrightarrow{\mathrm{s}} q b q a q.
\end{equation}
Since the nets $q a_n q, q
b_n q$ are uniformly bounded and the product on uniformly bounded sets is
strongly continuous \cite{Blackadar,TakesakiI},
\begin{equation}
    \begin{aligned}
        \t_q( (q a q ) (q b q)) &= \bra{\W}  (q a q ) (q b q)\ket{\W} \\
                                & = \lim_k\bra{\W}  (q a_k
                                q ) (q b_k q)\ket{\W} \\
                                & = \lim_k\bra{\W}  (q b_k
                                q ) (q a_k q)\ket{\W} \\
                                &= \bra{\W}  (q b q ) (q a q)\ket{\W}
    \end{aligned}
\end{equation}
where we used strong convergence of the nets in the second line which proves
cyclicity on $q M_\G q$. Since $\t_q$ is a restriction of a vector functional
to $qM_\G q$ it is normal. Since $q$ is the identity on $q\cH_\w$, we have that $\tau_q(q) =\bra{\W} q \ket{\W} = \langle \W | \W \rangle = 1$ by Lemma \ref{lem:q_exists}.
This implies that $\t_q$ defines a normal state on the von Neumann algebra $qM_\G q$. 
\end{proof}

\subsection{Faithfulness and finiteness}
Now that we established that the state $\t_q$ is tracial for $q M_\G q$, we
want to use it to restrict the full algebra $M_\G$. The central decomposition
of $M_\G$ asserts that we can decompose our Hilbert space into an integral
\begin{equation}
    \cH_\w \simeq \int_X \cH_x d\m(x),
\end{equation}
where $X$ is a measurable space equipped with a measure $\m$, such that for
each measurable set $E \subset X$ one obtains a projection $z_E$ which is in
the center of $M_\G$ such that
\begin{equation}
    z_E \cH_\w \simeq \int_E \cH_x d\m(x).
\end{equation}
Similarly one has a decomposition 
\begin{equation}
    M_\G \simeq \int_X M_x d\m(x),
\end{equation}
where $M_x \subset \cB(\cH_x)$ is a factor almost surely. Our goal is to prove
that none of the $M_x$ can have type III. To do so, we will prove that every
one of the subalgebras appearing in the central decomposition has a finite
projection, which arises from our $q$, which cannot happen for type III
algebras \footnote{They are defined as not having finite projections}.
The finiteness of a projection means that if there exists a partial isometry $v$
with
\begin{equation}
    v^\dagger v = q, \quad v v^\dagger \leq q,
\end{equation}
then $v v^\dagger = q$. A simple way to see finiteness is the existence of a
faithful tracial state $\t$, since then
\begin{equation}
    \t(q) = \t(v^\dagger v) = \t(v v^\dagger) \Rightarrow \t(q-v v^\dagger) =
    0.
\end{equation}
If $\t$ is faithful, i.e., for $x \geq 0$, $\t(x) = 0$ iff $x = 0$, this implies $q = v v^\dagger$.
We therefore need
\begin{proposition}[Faithfulness of the corner]\label{prop:faithful}
$\tau_q$ is faithful on $qM_\G q$.
\end{proposition}
\begin{proof}
Let $x\in q M_\G q$ satisfy
\begin{equation}
  \tau_q(x^\dagger x)=0.
  \label{eq:faithfulness-zero}
\end{equation}
We want to show that $x = 0$. We will do so, by demonstrating that $x \ket{\x}
= 0$ for a dense set of vectors $\ket{\x}$. This dense set 
will consist of vectors of the form
\begin{equation}\label{eq:dense_states}
    \ket{\x_{a,b}} = \pi_\w(a) \pi_\w(b) \ket{\W},
\end{equation}
 with $a \in \cA_\G, b \in \cA_{\G^c}$, which form a norm-dense set of $\cA$
 and since we consider the GNS Hilbert space of $\cA$ with respect to $\w$, these vectors form a
 dense set in $\cH_\w$.
For $y\in qM_\G q$ we have 
\begin{align}\label{eq:xyOmega_zero}
    \norm{xy\Omega}^2
  &=\tau_q(y^\dagger x^\dagger xy)\\
  \nonumber
  &=\tau_q(x^\dagger xy y^\dagger)\\
   \nonumber
  &=\tau_q\big((x^\dagger x)^{1/2}yy^\dagger(x^\dagger x)^{1/2}\big)\\
   \nonumber
  &\le \|y\|^2\tau_q(x^\dagger x)=0.
   \nonumber
\end{align}
In the second line we have used the definition of the state $\t_q$, in the third we
used traciality of $\t_q$ on $q M_\G q$. In the fourth line we used that every
positive operator has a square root together with traciality. In the last
line we used that 
\begin{equation}
    \norm{y}^2 \id \geq y y^\dagger,
\end{equation}
which follows from the definition of the operator norm. Conjugating the
positive operator $ O_y = \norm{y}^2 \id - y y^\dagger$ by $(x^\dagger x)^{1/2}$ again leads to a
positive operator. Since $\t_q$ is the restriction of a vector functional, it
is positive, which applied to the positivity of
$(x^\dagger x)^{1/2}O_y(x^\dagger x)^{1/2}$ implies the last line.
We thus have found that for every $y\in qM_\G q$
\begin{equation}\label{eq:all_xyOmega_zero}
  xy\ket{\W}=0.
\end{equation}
Now for general $a\in \cA_\G$, we note that $x=xq$, since $x \in q M_\G q$
and $q\ket{\W}=\ket{\W}$ by Lemma \ref{lem:q_exists}. This implies that
\begin{equation}\label{eq:x_a_zero}
  x\pi_\w(a)\ket{\W}
  =x q\pi_\w(a) q \ket{\W}=0.
\end{equation}
Since operators in $\cA_\G$ commute with operators in $\cA_{\G^c}$ and $\pi_\w$
is a representation of $\cA$, we have that
\begin{equation}
  M_\G
  =\pi_\omega(\cA_\G)''
  \subseteq
  \pi_\omega(\cA_{\G^c})'.
  \label{eq:M-commutes-complement}
\end{equation}
Since $x \in q M_\G q \subset M_\G$, it commutes with
$\pi_\omega( \cA_{\G^c})$.  For $a\in \cA_\G$ and $b\in \cA_{\G^c} $ we
therefore have on the dense set of vectors from Eq.~\eqref{eq:dense_states}
\begin{equation} \label{eq:x_global_annihilation}
  x\pi_\omega(a)\pi_\omega(b)\ket{\W}
  =\pi_\omega(b)x\pi_\omega(a)\ket{\W} = 0,
\end{equation}
which implies $x = 0$ and thus proves faithfulness.
\end{proof}
Note that we have only proved faithfulness of $\t_q$ as a functional on $qM_\G
q$, not of $\w$ $M_\G$. The discussion preceding Proposition
\ref{prop:faithful} needed faithfulness on the algebra to prove finiteness of
the projection. In the current situation we can still prove
\begin{corollary}[Finiteness of the limiting projection]\label{cor:q_finite}
The projection $q$ is finite in $M_\G$.
\end{corollary}

\begin{proof}
Assume the existence of a partial isometry $v\in M_\G$ with
\begin{equation}\label{eq:v_leq_q}
  v^\dagger v=q,
  \qquad
  vv^\dagger \leq q.
\end{equation}
The fact that $v$ is a partial isometry, together with the inequality in~\eqref{eq:v_leq_q}, implies that the range
of $v$ is contained in the range of $q$, on which $q$ acts as the identity. 
This implies that $v = qv$. 
By the same logic, the support of $v$ is contained in the support of $q$, so
that $v = vq$. Therefore $v = qvq$, which implies $v \in q M_\G q$.
Now we can apply the argument preceding Proposition \ref{prop:faithful}.
Traciality of $\t$ implies
\begin{equation}
  \tau_q(vv^*)=\tau_q(v^*v)=\tau_q(q)=1.
\end{equation}
Therefore $\tau_q(q-vv^*)=0$. Faithfulness of $\t_q$ and $q-vv^*\ge 0$ then
imply $vv^*=q$.  Thus $q$ is finite.
\end{proof}
The above proof verbatim also proves
\begin{lemma}[Finiteness is inherited from a corner]\label{lem:finite_corner_finite_global}
    Let $M$ be a von Neumann algebra and $q \in M$ a projection. Let $\t_q$ be a
    faithful tracial functional on $q M q$. Then $q$ is finite with respect to $M$.
\end{lemma}

\subsection{Fullness and exclusion of type III}
We now can move on to prove the main theorem. In the end, we want to
prove that for every central projection $z \in \cZ(M_\G)$, the algebra $zM_\G$ contains a nonzero finite projection. The projection we pick will be $zq$
and we therefore first need to show that, for every central projection $zq$ is
nonzero. This goes under the name of fullness.
The central support $c_{M_\G}(q)$ is the smallest central projection in
$M_\G$ dominating $q$ \cite[III.1.1.5]{Blackadar}.
The projection $q$ is called \emph{full} if its central support is the identity.
For a full projection $q$, if one has any nonzero central projection $z$, then $zq$ is
nonzero. If it were zero, then we would have that $q = (1-z)q$. For this to be
true, the
subspace associated to the projection $(1-z)$ must contain the subspace associated
to $q$, which is exactly
\begin{equation}
    q \leq (1-z),
\end{equation}
which contradicts fullness, since $(1-z)$ is central and dominates $q$.
Similarly, if $zq \neq 0$ for all central projections $z$, then there cannot be
a dominating projection $q \leq p$, since this implies that $(1-p)q = 0$, which
contradicts that $zq \neq 0$ for all central projections. 

Therefore, to claim that $zq$ is a nonzero finite projection, we need
\begin{proposition}[Fullness of the limiting projection]\label{prop:fullness}
The projection $q$ is full in $M_\G$.
\end{proposition}

\begin{proof}
We proceed by contradiction. Assume $q$ is not full. Then there exists a nonzero
$z\in\mathcal Z(M_\G)$ that satisfies
\begin{equation}\label{eq:zq_vanishes}
  zq=0.
\end{equation}
Since $q\ket{\W}=\ket{\W}$ by Lemma \ref{lem:q_exists},
\begin{equation}\label{eq:zW_vanishes}
  z\ket{\W}=z q \ket{\W}=0.
\end{equation}
Since $z \in \cZ(M_\G)$, it commutes with both $\pi_\omega( \cA_\G)\subset
M_\G$ and $\pi_\omega(\cA_{\G^c})\subset M'_\G$.  Therefore
\begin{equation}
  z\pi_\w(a)\pi_\w(b)\ket{\W}
  =\pi_\w(a)\pi_\w(b)z\ket{\W}=0
\end{equation}
for all $a\in \cA_\G$ and $b\in \cA_{\G^c}$. As described in the proof of
Proposition \ref{prop:faithful}, these vectors form a dense set. Thus $z = 0$
and there exist no nonzero central projection that is orthogonal to $q$. As
explained above the statement of the proposition, this implies that $q$ is
full. 
\end{proof}
We are now ready to prove our main theorem:
\begin{theorem}[Finite projections exist in every central sector]\label{thm:finite_projections_exist}
    Let $\cS$ be a consistent stabilizer system on a countable lattice, $\G
    \subset \cL$ a subregion, $\w$ the state asserted by Proposition
    \ref{prop:define_state} and $M_\G$ be the von Neumann algebra associated to
    $\G$ in the GNS Hilbert space $\cH_\w$. Then, for every nonzero central
    projection $z \in \cZ(M_\G)$,
    the von Neumann algebra $zM_\G$ contains a finite projection.
\end{theorem}

\begin{proof}
Let $z\in\mathcal Z(M_\G)$ be a nonzero central projection.  By
Proposition \ref{prop:fullness}, $q$ is full, which implies
\begin{equation}
  zq\ne0,
\end{equation}
as explained in the discussion preceding Proposition \ref{prop:fullness}.
Because $z$ is central, $zq$ is a projection
\begin{eqnarray}
        (zq)^\dagger &=& q^\dagger z^\dagger = q z = z q\\
        (zq)^2 &=& zq zq = z^2 q^2 = zq,
\end{eqnarray}
where we have used that $q,z$ are individually projections and by centrality $[q,z]
= 0$. The analogous calculation proves that $(1-z)q$ is a projection. Since
projections are positive operators, we have
\begin{equation}
 zq\le q.   
\end{equation}
  This implies that $(zq) M_\G(zq) \subset qM_\G q$. By Propositions
  \ref{prop:faithful} and \ref{prop:normal-trace} $\t_q$ is a normal faithful positive tracial
  state on $q M_\G q$. The restriction of $\tau_q$ to the subalgebra
  $(zq) M_\G(zq)$ is thus a faithful positive tracial functional, so $zq$ is
  finite with respect to $(zq) M_\G (zq)$.
  Since $(zq) M_\G (zq) = (zq) (z M_\G) (zq) \subset z M_\G$ and $zq \in zM_\G$, $zq$ is also finite with respect to $z
  M_\G$ by Lemma \ref{lem:finite_corner_finite_global}.
  Therefore every nonzero central summand $z M_\G$ contains a nonzero
  finite projection which proves the assertion.
\end{proof}

\section{Stabilizer systems with short-range order}\label{app:short_range}
This appendix treats the special case of short-range ordered systems, where the stabilizers have a simpler structure,
which allows the decomposition of the state into individual clusters. This allows more control over the resulting algebras.
In the following we assume that all stabilizer systems have \emph{bounded
support}, i.e., there is a $N > 0$ such that 
\begin{equation}
    \mathrm{diag}(\operatorname{supp}(s)) < N,\quad s \in \cS,
\end{equation}
where 
\begin{equation}
    \mathrm{diag}(M) = \inf\{r>0 : M\subset B_r(x)\text{ for some }x\in M\},
\end{equation}
is the size of the smallest ball into which $M$ fits.

A convenient way \cite{Calderbank:1996aj} to talk about stabilizer systems, or Pauli operators in general, is by thinking of Pauli
operators as represented by tuples of bits $(x_i,z_i)$ for each site $i \in
\cL$, via the map
\begin{equation}
    P(v) = \prod_{i \in \cL} X_i^{x_i} Z_i^{z_i},\qquad v=(x\mid z)\in V_{\mathrm{fin}}
\end{equation}
For a finite region $\Lambda$, define
\begin{equation}
    V_\Lambda:=\mathbb{Z}_2^\Lambda\oplus\mathbb{Z}_2^\Lambda,
\end{equation}
the vector space of $\mathbb{Z}_2$-valued maps on $\Lambda$.
An element is written $v=(x\mid z)$, where $x,z:\Lambda\to\mathbb{Z}_2$.  Its
support is
\begin{equation}
    \operatorname{supp}(v):=\{a\in\Lambda:(x_a,z_a)\ne(0,0)\}.
\end{equation}
For an infinite region we use 
\begin{equation}
    V_{\mathrm{fin}}:=\{v:\cL\to\mathbb{F}_2^2:\operatorname{supp}(v)\text{ is finite}\}.
\end{equation}
This is the vector space of finitely supported labels.
We identify $V_\L \subset V_{\mathrm{fin}}$ by extending labels by zero outside
$\L$.

Then, there is a phase function
$\sigma:V_{\mathrm{fin}}\times V_{\mathrm{fin}}\to U(1)$ such that
\begin{equation}\label{eq:pauli_product}
  P(v)P(w)=\sigma(v,w)P(v+w).
\end{equation}
In the notation above we have for $v = (x_a,z_a), w = (x'_a,z'_a)$ that 
\begin{equation}\label{eq:pauli_phase}
    \s(v,w) = (-1)^{\sum_a z_a x'_a}.
\end{equation}
These Pauli operators satisfy 
\begin{equation}\label{eq:adjoint_pauli}
    P(v)^\dagger = \s(v,v) P(v).
\end{equation}
This implies that 
\begin{equation}\label{eq:self_adjoint_pauli}
    Q(v) = i^{\sum_a z_a x_a} P(v) 
\end{equation}
is self-adjoint.
For the commutation relations it is useful to use
\begin{definition}[Symplectic form and symplectic complement]
For $v=(x\mid z)$ and $w=(x'\mid z')$ in $V_{\mathrm{fin}}$, define
    \begin{equation}\label{def:symplectic_pauli_form}
    \langle v,w\rangle :=  \sum_{a \in \cL}(x_a z'_a+z_a x'_a) \pmod 2.
    \end{equation}
    This defines a non-degenerate and alternating bilinear form on
    $V_{\mathrm{fin}}$. Since both vectors have finite support, the sum is
    finite. For a subspace
$E\subset V_{\mathrm{fin}}$, define
\begin{equation}
    E^\perp:=\{v\in V_{\mathrm{fin}}:\langle v,e\rangle=0\text{ for all }e\in E\}.
\end{equation}
\end{definition}
The choice is normalized by \(P(0)=\id\).    

One then has   
\begin{equation}\label{eq:pauli_commutation}
  P(v)P(w)=(-1)^{\langle v,w\rangle}P(w)P(v).
  \end{equation}

We see that we can associate, up to a phase, to each Pauli operator a vector $v
\in V$. In particular, Eq.~\eqref{eq:pauli_product} demonstrates that the group
generated by a set of Paulis is determined by the subspace $E$ spanned by their
associated vectors $v \in V$ and its commutant is the group generated by the
Paulis associated to $E^\perp$.
Given any set $\cO$ of operators in $\cA_{\text{loc}}$ and a finite subset $\L
\Subset \cL$, we define the local set $\cO_\L$
via 
\begin{equation}
    \cO_\L = \cO \cap \cA_\L,
\end{equation}
where we pad operators with identities outside of $\L$ using
\eqref{eq:lattice_embedding}. This for example defines the \emph{local
stabilizer group} $\cG^{\cS}_{\L}$. It should be noted that this set can be
larger than the group generated by just the local stabilizers, because it can
happen that stabilizers that are supported outside of the region $\L$ can have
a product, that is supported only in $\L$. This definition of the local
stabilizer group is necessary in Proposition~\ref{prop:define_state} to have global
consistency of the state.

\subsection{Isotropy}
\begin{definition}[Isotropic and maximally isotropic subspaces]
    Let $V$ be a vector space over $F$ and $\langle\cdot,\cdot \rangle$ an
    alternating nondegenerate bilinear form $V \times V
    \rightarrow F$. A subspace $M \subset V$ is isotropic, if 
    \begin{equation}
         \langle m,m'\rangle = 0, \quad m,m' \in M.
    \end{equation}
    $M$ is called maximally isotropic if it is not a proper subspace of any other isotropic subspace, i.e., if $N$ is isotropic and 
    \begin{equation}
        M \subseteq N \subseteq V,
    \end{equation}
    then $N=M$. For maximally isotropic subspaces of
    finite-dimensional vector spaces one has $2 \dim(M) = \dim(V)$. 
\end{definition}

\begin{definition}[Symplectic bases and hyperbolic pairs]
    Let $V$ be a vector space with alternating nondegenerate bilinear form
    $\langle.,.\rangle$. A
    basis of $V$ is called a symplectic basis if it can be written as a set of tuples $\{z_i,x_i, i \in I \}$ with 
    \begin{equation} 
                \langle z_i,z_j\rangle=\langle x_i,x_j\rangle=0,
                  \quad
                    \langle z_i,x_j\rangle=\delta_{i,j}.
    \end{equation}
    The tuples $(z_i,x_i)$ are called hyperbolic pairs.
\end{definition}
By starting out with an isotropic subspace $U \subset U^\perp$ with basis $u_i$, Lemma 7.5 in Ref.~\cite{Taylor1992}
implies that there exists an isotropic subspace $U'$ such that $V = U^\perp
\oplus U'$ and there is a unique basis $u'_i$ of $U'$ such that $(u_i,u'_i)$
are hyperbolic pairs $\langle u_i,u'_j \rangle = \d_{i,j}$. The
discussion in the section Symplectic Bases in Ref.~\cite{Taylor1992} then extends
this to a symplectic basis of $V$.
The stabilizer group restricted to a subregion forms an isotropic subspace. An adapted basis corresponds to Pauli operators that generate the group. However, if $z_i$ span the local stabilizer group, there is no guarantee that their hyperbolic partner $x_i$ is also local. If however, all stabilizers fall into clusters, such as illustrated in Figure \ref{fig:stabilizers-short-range-order}, then the Pauli group definitely is spanned by local pieces and one can find an adapted basis, that is associated to $W_\G$. We lift this observation to a definition of short-range order:

\begin{definition}[Short-range order]\label{def:short_range_order}
    Let $\G \subset \cL$ be a subregion and $W_\G \subset V_\G$ the subspace associated to the stabilizer group. We say that $\cS$ induces short-range order in $\G$ if there exists 
    a symplectic basis $\{z_i,x_i : i\in I\}$ of $V_\Gamma$, with
    $|I|=|\Gamma|$, and a subset $P\subset I$
        such that
    \begin{equation}
        W_\Gamma=\operatorname{span}_{\mathbb F_2}\{z_i : i\in P\}.
    \end{equation}
    and every hyperbolic pair $(x_i,z_i)$ is supported on a finite region.
\end{definition}
This definition guarantees that we can build $W_\G$ from local pieces and there is no long-range structure in the state. This condition is implied by the existence of a constant depth local Clifford circuit that prepares the stabilizer system. The existence of a constant depth local circuit that prepares a vector from a product vector is the conventional definition of short-range entanglement \cite{Chen:2010gda}, which motivates our choice to call the above systems short-range ordered.

\subsection{Local algebras in short-range ordered stabilizer systems}
    
For a general stabilizer system, we distinguish two cases: those with long-range
order and without long-range order. A system with long-range order is given for
example on a lattice indexed by $\mathbb{Z}$ and stabilizers $S_n = Z_n
Z_{n+1}$. The code space defined by these stabilizers is spanned by the all-zero
or all-one state, so that the local density matrix for a finite system is 
\begin{equation}
    \r = \frac{1}{2} (\ket{0, \hdots,0}\bra{0,\hdots,0} + \ket{1, \hdots,
    1}\bra{ 1, \hdots, 1}).
\end{equation}
The GNS Hilbert space of the local algebra then becomes a direct sum
\begin{equation}
    \cH_\w = \cH_0 \oplus \cH_1,
\end{equation}
with the all-zero or all-one state being the respective vacuum of $\cH_0$ or $\cH_1$.
Because the all-zero state is unentangled, we have that 
\begin{equation}
    \cH_0 = \cH^-_{0} \otimes \cH^+_{0},
\end{equation}
where these are the Hilbert spaces that arise from finite excitations of the
qubits for $n<0$ and $n \geq 0$.
The local algebras for e.g. the half-chain $A_{\mathbb{N}}$, will become 
a direct sum of the form
\begin{equation}
    \pi_\w(\cA_{\mathbb{N}})'' \cong \big( \id \otimes \cB(\cH^+_0) \big) \oplus
    \big( \id \otimes \cB(\cH^+_1) \big).
\end{equation}
The two summands can be obtained by taking the strong limit
\begin{equation}
    Z_n \xrightarrow{\mathrm{s}} P.
\end{equation}
$P$ is then an operator that has all vectors in $\cH_0$
as eigenstates with eigenvalue $+1$ and all the vectors in $\cH_1$ as eigenvectors
with eigenvalue $-1$. The center of $\pi_\w(\cA_{\mathbb{N}})''$ is then generated by the
projections
\begin{equation}
    P_1 = \frac{1}{2}(\id + P),\quad P_2 = \frac{1}{2} (\id - P).
\end{equation}
Therefore in this case, the algebra generated by $\cA_{\mathbb{N}}$ is not a factor and has a nontrivial
center generated by $P_1$ and $P_2$.
From the stabilizers one can see that the algebra has a nontrivial
center, because one cannot choose a symplectic basis adapted to $W_{+}$ in
$V_{\mathrm{fin}}$. One way to see it is that the
isotropic space associated to the stabilizer system has vectors that are
supported on regions that are arbitrarily far apart, e.g. the vector associated
to
\begin{equation}
    \prod_{n=1}^{k} S_n = Z_1 Z_{k+1}
\end{equation}
The associated vector in $V_{\mathrm{fin}}$ is $v=(0\mid e_1+e_{k+1})$ but
each of the terms themselves is not individually an element of $W_\G$. We can
therefore not span $W_\G$ as a direct sum of local spaces. This is
what induces long-range order and which makes the algebra not a factor, because
the individual vectors on the code look very different at large scales.

We can now state the main result for infinite stabilizer systems without
long-range order:
    \begin{theorem}[Classification of short-range stabilizer algebras]\label{thm:short_range_algebra}
        Let $\cS$ be a consistent stabilizer system on a countable lattice $\cL$. Let $\w$ be the associated state on the quasi-local algebra $\cA$ with associated
        GNS Hilbert space $\cH_\w$. Let $\G \subset \cL$ be a subset of the lattice on which $\cS$ induces short-range order and $M_\G$ its
        associated von Neumann algebra in $\cH_\w$. Let $I,P$ be the index set labelling the associated hyperbolic basis as in Definition \ref{def:short_range_order}. Then 
        \begin{equation}
            M_\G \cong \cB(\cH_P)\bar\otimes\cR_{I\setminus P},
        \end{equation}
        where $\bar{\otimes}$ is the von Neumann tensor product, $\cH_P$ is a Hilbert space with dimension $2^{|P|}$ if $P$ is finite or separably infinite-dimensional if $|P|$ is infinite, and $\cR_{I\setminus P}$ is either a type I$_{2^{|I \setminus P| }}$ factor if $|I \setminus P|$ is finite or $\cR$ the hyperfinite type II$_1$ factor.
    \end{theorem}

    To prove Theorem \ref{thm:short_range_algebra}, we will be studying what the state looks like only in the region $\G$. This leads to one GNS representation of $\cA_\G$, which only contains vectors that arise from acting with operators located in $\G$ on the vacuum $\w$. We call this the \emph{local} representation $\pi_\G$.
Formally, this is a different representation than the GNS representation of the quasi-local algebra $\cA$ restricted to $\cA_\G$, which we denote by $\pi_{\w,\G}$, because the Hilbert space also contains vectors obtained by acting with operators supported outside of $\G$ on $\ket{\W}$. We call this the \emph{global} representation.
We will first prove the structure theorem \ref{thm:short_range_algebra} for the local representation.
To determine the structure of the local algebra of Theorem \ref{thm:short_range_algebra} in the global representation, which is what we are after, we need to show that these representations are equivalent. Thus we need:
\begin{lemma}[Quasi-equivalence of local and global GNS representations]\label{lem:equivalence-local-global}
    Let $\cS$ be a consistent stabilizer system on a lattice $\cL$, and let $\G \subset \cL$ be a region with local $C^\ast$-algebra $\cA_\G$. Let $\w$ be the state of Proposition~\ref{prop:define_state} on the quasi-local algebra $\cA$, $\pi_\G$ the GNS representation of $\cA_\G$ with respect to $\w$ and $\pi_{\w,\G}$ be the restriction of $\pi_\w$ to the local algebra $\cA_\G$. Then $\pi_{\w,\G}$ is quasi-equivalent to $\pi_\G$.
\end{lemma}
    
\begin{proof}
    According to Proposition \ref{prop:equivalence-representations} it suffices to show that the folia in both representations agree, i.e.,  
    \begin{equation}
        \cF_{\pi_{\w,\G}}(\cA_\G) = \cF_{\pi_\G}(\cA_\G).
    \end{equation}
    By Lemma~\ref{lem:linear-folium} we know that the closed invariant subspace associated to a representation is the norm closure of the linear span of the functionals arising from matrix elements. It is clear that all functionals $\w_{\xi,\eta}$ defined by vector states $\xi,\eta \in \cH_\G$ in the Hilbert space of the local representation can be represented as vector states in the global GNS Hilbert space, as the local GNS Hilbert space is a subspace of the global Hilbert space, so that 
	    $\cF_{\pi_{\w,\G}}(\cA_\G) \supset \cF_{\pi_\G}(\cA_\G)$.
	    Thus, the nontrivial step is to show the converse $\cF_{\pi_{\w,\G}}(\cA_\G) \subseteq \cF_{\pi_\G}(\cA_\G)$.
	    As shown in Lemma~\ref{lem:linear-folium},
	    $\cF_{\pi_{\w,\G}}(\cA_\G)$ is the norm closure of the span of
	    functionals of the form $\w^{\pi}_{\x,\h}$. We
	    will thus show that each such functional is an element of
        $\cF_{\pi_\G}(\cA_\G)$. We will show it for a dense set of vectors in
        $\cH_\w$ that arise from finite Pauli excitations of the vacuum $\w$.
        The passage from this dense set to arbitrary vectors follows
        by continuity of vector functionals in the Hilbert-space norm and is
        given at the end of this proof.
        So take two vectors $\psi_u$ and
	    $\psi_{u'}$, where
	    \begin{equation}
	        \ket{\psi_u} = P(u)\ket{\W},
	        \quad
	        \ket{\psi_{u'}} = P(u')\ket{\W},
	    \end{equation}
	    and $P(u)$ and $P(u')$ are Pauli operators associated to $u$ and
	    $u'$. Now consider the linear functional $\w^{\pi}_{\y_u,\y_{u'}}$.
        For any local Pauli $P(v) \in \cA_{\G}$ we have 
        \begin{align}
            \w^{\pi}_{\y_u,\y_{u'}}(P(v))
            &=
            \mel{\y_u}{\pi_{\w}(P(v))}{\y_{u'}} \notag\\
            &=
            \mel{\W}{\pi_\w(P(u))^\dagger \pi_\w(P(v))\pi_\w(P(u'))}{\W}
            \notag\\
            &=
            \s(u,v,u')\mel{\W}{\pi_\w(P(u+v+u'))}{\W}.
        \end{align}
        where $\s$ is a phase. We absorbed the adjoint on $P(u)$ into the phase
        $\s$ for notational convenience.
        We now want to show that this functional can be
        recovered just by acting with an operator supported in $\G$. First, if
        $u+v+u' \notin W$, then the expectation value vanishes. In the
        following assume that 
        \begin{equation}\label{ass:v-copy}            
        u + v + u' \in W.
        \end{equation}
        Then we have that $\w^{\pi}_{\y_u,\y_{u'}}(P(v)) \in \mathrm{U}(1)$ is
        a phase.
        Now we want to find an
        operator that creates this expectation value but acts only in $\G$. For
        this, choose a $v_0 \in V_\G$ with
        \begin{equation}\label{ass:v_0-copy}
            u + v_0 + u' \in W.
        \end{equation}
        If no such $v_0$ exists, then $u+v+u'\notin W$ for every local $v\in
        V_\G$. By the definition of $W$, $\w$ vanishes on such Pauli operators.
        Therefore locally, the functional is zero and already an element of
        $\cF_{\pi_\G}(\cA_\G)$. Hence we assume that such a $v_0$ exists.
        As we will see, we can use this $v_0$ to create our functional from
        vectors in $\cH_{\G}$.
        By assumption~\eqref{ass:v-copy}, the expectation value $\w(P(u+v+u'))
        \neq 0$ and equals a phase, since $\pi_\w(P(u+v+u'))\ket{\W} =
        e^{i\phi(u+v+u')}\ket{\W}$ with $\phi(u+v+u') \in \mathbb{R}$, since
        $P(u+v+u')$ stabilizes $\ket{\W}$ up to a phase by the definition of
        $W$. We can characterize such $v$ as follows. Since the stabilizer
        group is a group, products of group elements are also in the group. For
        their respective labels this means that
        \begin{equation}
            u+v+u'+u+v_0 +u' = v_0 + v \in W,
        \end{equation}
        where we used that $u+u = 0$ since $u$ is an element of a vector space
        over $\mathbb{F}_2$. We thus find that
        \begin{equation}
            u + v + u' \in W \Leftrightarrow  v + v_0 \in W_\G,
        \end{equation}
        where we used the definition $W_\G=W\cap V_\G$ and the fact that
        $v+v_0\in V_\G$.
        Therefore, after choosing $v_0$, we see that the only operators that have nonzero
        value under $\w_{\y_u, \y_{u'}}^{\pi}$ are associated to vectors of the
        form $v = v_0 + s, s \in W_\G$, where $s = v_0 + v$.
        On these elements, put
        \begin{equation}
            k:=u+v_0+u'\in W.
        \end{equation}
        We will have to keep track of the phases. For $w\in W$, let $\chi(w)\in
        \mathrm{U}(1)$ be the phase that relates $P(w)$ to the associated
        element of the stabilizer group, i.e.,
        \begin{equation}
            \chi(w)P(w)\ket{\W}=\ket{\W}.
        \end{equation}
        We then have that 
        \begin{align}
        \omega^\pi_{\psi_u,\psi_{u'}}(P(v_0+s))
        &=
        \sigma(u,u)^{-1}
        \sigma(u,v_0+s)
        \sigma(u+v_0+s,u')
        \omega(P(u+v_0+s+u'))\\
        &=
        \sigma(u,u)^{-1}
        \sigma(u,v_0)
        \sigma(u+v_0,u')
        \sigma(u,s)\sigma(s,u')
        \omega(P(k+s)).
        \end{align}
        Since $\chi(k)P(k)$ and $\chi(s)P(s)$ are stabilizers, Eq.~\eqref{eq:pauli_product} gives
        \begin{equation}
            \w(P(k+s))
            =
            \chi(k)^{-1}\s(k,s)^{-1}\w(P(s))
            =
            \chi(k)^{-1}(-1)^{z_k\cdot x_s}\w(P(s)).
        \end{equation}
        Hence
        \begin{equation}\label{eq:functional_global}
            \w_{\y_u,\y_{u'}}^{\pi}(P(v_0+s))
            =
            c\chi(k)^{-1}
            (-1)^{(z_u+z_k)\cdot x_s+z_s\cdot x_{u'}}
            \w(P(s)),
        \end{equation}
        where we defined the $s$-independent constant
        \begin{equation}
            c:=\s(u,u)^{-1}\s(u,v_0)\s(u+v_0,u').
        \end{equation}
        Since $s\in V_\G$, only the restrictions
        of $z_u+z_k$ and $x_{u'}$ to $\G$ contribute to the phase. We then set
        \begin{equation}
            r:=
            \left(
                x_{u'}\vert_\G+x_{v_0}
                \,\middle|\,
                (z_u+z_k)\vert_\G
            \right)
            \in V_\G.
        \end{equation}
        Now consider the local functional
        \begin{equation}\label{eq:local_functional}
            \begin{aligned}
            \bra{\pi_\G(P(r))\W_\G}
            \pi_\G(P(v_0+s))
            \pi_\G(P(v_0+r))
            \ket{\W_\G}
            &=
            \s(r,r)
            \bra{\W_\G}
            \pi_\G(P(r))
            \pi_\G(P(v_0+s))
            \pi_\G(P(v_0+r))
            \ket{\W_\G}\\
            &=
            \s(r,r)(-1)^{(z_{v_0}+z_s)\cdot(x_{v_0}+x_r)}
            \bra{\W_\G}
            \pi_\G(P(r))
            \pi_\G(P(s+r))
            \ket{\W_\G}\\
            &=
            \s(r,r)(-1)^{z_r\cdot x_s+z_r\cdot x_r+
            z_{v_0}\cdot x_{v_0}+z_s\cdot x_{v_0}+
            z_{v_0}\cdot x_r+z_s\cdot x_r}
            \w(P(s))\\
            &=
            c'(-1)^{z_r\cdot x_s+z_s\cdot x_{v_0}+z_s\cdot x_r}
            \w(P(s))
            \end{aligned}
        \end{equation}
        where we repeatedly used Eqs.~\eqref{eq:pauli_phase} and
       ~\eqref{eq:pauli_product} and defined the $s$-independent phase
        \begin{equation}
            c':=\s(r,r)(-1)^{z_r\cdot x_r+z_{v_0}\cdot x_{v_0}+
            z_{v_0}\cdot x_r}.
        \end{equation}
        Comparing this with the global functional in Eq.
       ~\eqref{eq:functional_global}, we see that the two functionals differ only by the
        $s$-independent phase $c\chi(k)^{-1}(c')^{-1}$ and both vanish if $s \notin W_\G$. 

        Note that we used the local representation $\pi_\G$. The above
        expression is well defined since by construction $r,v_0 \in V_\G$.
        By multiplying the functional by this phase, we thus have proved
        their coincidence on all Pauli operators supported in $\G$. 
        Since the linear span of local Pauli operators is norm dense in $\cA_\G$, both bounded linear functionals agree on all of $\cA_\G$. This proves that
        $\w^\pi_{\xi,\eta}\in\cF_{\pi_\G}(\cA_\G)$ for all
        $\xi,\eta\in\cD$, where
        \begin{equation}
            \cD:=\operatorname{span}\{P(u)\ket{\W}:u\in V_{\mathrm{fin}}\}.
        \end{equation}
        This subspace is dense in $\cH_\w$. Let
        $\xi_n,\eta_n\in\cD$ converge respectively to
        $\xi,\eta\in\cH_\w$. Then
        \begin{equation}
            \norm{\w^\pi_{\xi,\eta}-\w^\pi_{\xi_n,\eta_n}}
            \leq
            \norm{\xi-\xi_n}\norm{\eta}
            +
            \norm{\xi_n}\norm{\eta-\eta_n}
            \to 0.
        \end{equation}
        Since $\cF_{\pi_\G}(\cA_\G)$ is norm closed, this gives
        $\w^\pi_{\xi,\eta}\in\cF_{\pi_\G}(\cA_\G)$ for all
        $\xi,\eta\in\cH_\w$.
        Therefore the states are
        vectors in the local representation. Hence
	        $\w^{\pi}_{\y_u,\y_{u'}}\in\cF_{\pi_\G}(\cA_\G)$.
           Due to linearity, the span of such states is also in
           $\cF_{\pi_\G}(\cA_\G)$ and since $\cF_{\pi_\G}(\cA_\G)$ is closed we
           have that
           \begin{equation}
               \cF_{\pi_{\w,\G}}(\cA_\G) \subseteq \cF_{\pi_\G}(\cA_\G)
           \end{equation}
           which we wanted to show.
        \end{proof}

    \begin{proof}[Proof of Theorem \ref{thm:short_range_algebra}]
    Given a consistent stabilizer system $\cS$, we have defined a state
    $\w$ on
    the quasi-local algebra $\cA$ which induces through its
    GNS-construction a Hilbert space $\cH_\w$. Given a subsystem $\G
    \subset \cL$ of the lattice, we now want to determine the structure of the local algebras.
    Because the stabilizers commute, they define an isotropic subspace $W
    \subset V_{\text{loc}}$, whose elements label all finite products of
    stabilizers. The intersection
    \begin{equation}
        W_\G = W \cap V_{\G}.
    \end{equation}
    By the assumption of short-range order, we can find a symplectic basis $\{z_i, x_i : i \in I \}$, where $I$ has the
    same cardinality as the sites in $\G$, consisting of hyperbolic pairs of $V_\G$
    which contains a subset labelled by $P \subset I$ with
    \begin{equation}
        W_\G = \mathrm{span}_{\mathbb{F}_2} \{z_i, i \in P \}.
    \end{equation}
    Note that $z_i,x_i$ are vectors in $V_\G$ that are not just vectors
    with a single nonzero entry. 
    The existence of short-range order allows us to put the subsystem in $\G$ into normal form \cite{Calderbank:1996aj} and map it into a tensor product representation by constructing an $\ast$-isomorphism $\a$ between the local $C^\ast$ algebra $\cA_\G$ and the quasi-local $C^\ast$-algebra $\cA_I$ of Pauli operators $\cP_I$ on a lattice of cardinality $I$, which again is defined as the norm completion of the union of local algebras, i.e.,
    \begin{equation}
        \cA_I = \overline{\bigcup_{F \Subset I} \bigotimes_{i \in F} M_2(\mathbb{C})}^{\norm{.}}.
    \end{equation}
    Here again, we pad operators with identities.
    On single Pauli operators we define
    \begin{equation}\label{eq:define_alpha}
        \a(X_i) = Q(x_i),\quad \a(Z_i) = \ve_i Q(z_i),
    \end{equation}
    where $Q$ is the self-adjoint operator of Eq.~\eqref{eq:self_adjoint_pauli} and 
    \begin{equation}
        \ve_i = \begin{cases}
        1 \quad \text{if } i \notin P\ \text{or } i \in P \text{ and }Q(z_i) \in \cG^{\cS}_\G,\\
        -1 \quad \text{else.} 
        \end{cases}
    \end{equation}
    The prefactor takes care of the case if $Q(v)$ or $-Q(v)$ is in the stabilizer group.
    To extend it to the full Pauli we have to define it on products. We do this by defining $\a$ on operators of the form
    \begin{equation}
        W(a,b) = \prod_{i} X^{a_i}Z^{b_i},
    \end{equation}
    with $a,b$ bit-strings with finite support. On these operators we have
    \begin{equation}
        \a(W(a,b)) = \prod_{i} \a(X^{a_i})\a(Z^{b_i}),
    \end{equation}
    and we extend $\a$ by linearity to scalar multiples and sums. This defines it on the norm-dense set of the $C^\ast$-algebra $\cA_I$. We now will show that $\a$ can be extended to an $\ast$-isomorphism between $\cA_I$ and $\cA_\G$.
    Since the $x_i,z_i$ form a symplectic basis the properties, self-adjointness, commutation relation, product relations of Pauli operators are preserved. 
    Then we prove it is injective. For each finite subset $F \Subset I$, the restriction of $\a$ to $\cA_F$ maps two distinct Pauli operators $W(a_1,b_1), W(a_2,b_2)$ in $\cA_F$ to two distinct Pauli operators in $\cA_\G$, because 
    \begin{equation}
        \a(W(a,b)) \propto P(\sum_i a_i x_i+b_i z_i),
    \end{equation}
    so if $(a_1,b_1) \neq (a_2,b_2)$ the images will have different binary labels and thus be different operators, since Paulis are linearly independent. 
    
    Since they form a basis, this implies that $\a$ is injective. This means that $\a$ defines an injective $\ast$-homomorphism 
    \begin{equation}
        \alpha_F: \bigotimes_{i \in F} M_2(\mathbb{C}) \rightarrow \cA_\G
    \end{equation}
    An injective $\ast$-homomorphism is isometric \cite{Blackadar}. 
    The restrictions are compatible under inclusions $F \subset F'$ since they arise from the same map $\a$. So $\a$ defines an isometric $\ast$-homomorphism on a norm-dense set of $\cA_I$ and therefore extends uniquely to an isometric $\ast$-homomorphism on $\cA_I$.
    Since the $(x_i,z_i)$ form a basis of $V_\G$, every Pauli operator in $\cA_\G$ is in the image of $\a$, as it can be written as a product of the operators $\a(X_i),\a(Z_i)$. This shows that a norm-dense set of $\cA_\G$ is in the image of $\a$ and since $\a$ is isometric and $\cA_I$ is norm-closed, its image contains the closure of this norm-dense set, which is $\cA_\G$. Therefore $\a$ is surjective, which proves that it is a $\ast$-isomorphism $\a:\cA_I \rightarrow \cA_\G$.
    
    This in turn defines a state on $\cA_I$ via
    \begin{equation}
        \tilde{\w} = \w \circ \a.
    \end{equation}
    For every local product $F \Subset I$ one has that on operators in $\cA_F$
    embedded in $\cP_I$ that the local density matrix is
    \begin{equation}\label{eq:tilde_omega}
        \tilde{\w}\vert_{\cA_F} = \bigotimes_{i \in (F \cap P)} \ket{0}\bra{0} \otimes \bigotimes_{j
        \in F \setminus P}
        \frac{\id_j}{2}.
    \end{equation}
    This can be inferred by checking it on finite products of Pauli operators, which form a basis of
    the operators. For the simple operators $W(a,b)$ we have that if any $a_i,b_i \neq 0$ for $i \notin P$ or $a_j \neq 0$ for $j \in P$,
    then $\tilde{\w}(W(a,b)) = 0$, as $\alpha(O)$ is proportional to an operator $P(v)$ with $v \notin W_\G$ by the definition of the symplectic basis. If $a_i = 0$ for all $i\in I$ and $b_i \neq 0$ only for $i \in P$  
    then the operator $\a(W(a,b))$ will by the definition Eq.~\eqref{eq:define_alpha} be a product of stabilizers, which have expectation values $1$ in $\w$. One can check that the state
   ~\eqref{eq:tilde_omega} has the same expectation value. Since the states
    agree on a basis of operators, they are the same states. Thus the state
    on $\G$ can be represented by a density matrix which is a tensor product of
    pure states or maximally mixed states. We can then form the GNS
    Hilbert space $\cH_{\tilde{\w}}$ with representation $\pi_{\tilde{\w}}$ of $\cA_I$, which is an infinite tensor
    product that we can decompose into a tensor product 
    \begin{equation}
        \cH_{\tilde{\w}} = \cK \otimes \cK'.
    \end{equation}
    If $|P| < \infty$, $\cK$ is the Hilbert space of $|P|$ qubits and if $|P| = \infty$, $\cK$ is a separable infinite-dimensional Hilbert space.
    $\cK'$ is the GNS Hilbert space associated to the maximally mixed state on a $C^\ast$-algebra of Pauli operators on $|I\setminus P|$ qubits, which is separable infinite-dimensional if $|I \setminus P| = \infty$ and otherwise has dimension $4^{|I\setminus P|}$.
    We therefore arrive at a tensor product of Araki-Woods factors \cite{ArakiWoods1968} on
    \begin{equation}
        \pi_{\tilde{\w}}(\cA_I)'' = \cB(\cK) \bar{\otimes} \cR_{I\setminus P}.
    \end{equation}
    As alluded to in the theorem statement, the exact types depend on whether we have finitely or infinitely many maximally
    entangled or pure factors in the state $\tilde{\w}$. If $|P| = \infty$ there are
    infinitely many pure pieces $\ket{0}$, then $\cB(\cK)$ is a type I$_\infty$ factor,
    otherwise it is type I$_{2^{|P|}}$.
    This coincides with the dimension of $W_\G$. If we have infinitely many
    maximally entangled pieces, then $\cR$ is the hyperfinite type II$_1$
    factor, otherwise it is again a finite matrix algebra of type I$_{2^{|I\setminus P|}}$. The $\ast$-isomorphism $\alpha$ provides us with an isomorphism 
    \begin{equation}
        \pi_{\tilde{\w}} \cong \pi_{\G} \circ  \alpha.
    \end{equation}
    Therefore 
    \begin{equation}
        \pi_\G(\cA_\G)'' \cong \cB(\cK) \bar{\otimes} \cR_{I\setminus P}.
    \end{equation}
    By Lemma \ref{lem:equivalence-local-global} this representation is quasi-equivalent to the restriction of $\pi_\w$ to $\cA_\G$. Thus it follows that 
    \begin{equation}
        M_\G \cong \cB(\cK) \bar{\otimes} \cR_{I\setminus P}.
    \end{equation}
    which we wanted to prove.
\end{proof}

\section{Probabilistic preliminaries}\label{app:concentration}
In the later proof we need the notion of \emph{concentration function} of a
random variable 
\begin{definition}[Concentration function]\label{def:concentration_function}
    Let $X$ be a real-valued random variable. Its concentration function
    $Q_X(t)$ is defined by 
    \begin{equation}
        Q_X(t) := \sup_{a \in \mathbb{R}} \mathbb{P}\{ |X-a| \leq t \},\ t \geq 0.
    \end{equation}
\end{definition}
An upper bound for $Q_X(t)$ is an anti-concentration inequality. The following inequality first proved by Kolmogorov
\cite{Kolmogorov1958} and later refined by
Rogozin in Ref.~\cite{Rogozin:1961} is relevant for us.
\begin{theorem}[Kolmogorov--Rogozin inequality, common scale]
    Let $X_1, \hdots, X_n$ be independent random variables and $S =
    \sum_{i=1}^{n} X_i$. Then there exists a constant $C$ such that for any $t > 0$ and any $0< \ell < t$,
    we have 
    \begin{equation}
        Q_S(t) \leq  \frac{C t}{\ell \sqrt{\sum_{i=1}^{n}(1-Q_{X_i}(\ell))}}.
    \end{equation}
\end{theorem}
\begin{theorem}[Kolmogorov--Rogozin inequality, variable scales]
    In the preceding setup, there exists a constant $C$ such that for any $t >
    0$ and any $0< \ell_i \leq t,\ i \leq n$,
    one has 
    \begin{equation}
        Q_S(t) \leq  \frac{C t}{ \sqrt{\sum_{i=1}^{n}\ell_i^2(1-Q_{X_i}(\ell_i))}}.
    \end{equation}
\end{theorem}

\subsection{Concentration for rescaled Bernoulli random variables}\label{app:Bernoulli}
Now consider a random variable of the form $X = c Y$ with $Y \sim
\mathrm{Bernoulli}(p)$ and $c > 0$ some constant. This variable takes either
the value $0$ with probability $(1-p)$ or the value $c$ with probability $p$.
Its concentration function can be computed as
follows: For $2t < c$, the interval $[a-t,a+t]$ can contain at most one of the
two values $\{0,c\}$, thus we have 
\begin{equation}\label{eq:conc_b_1}
    Q_X(t) = \max\{p,1-p\},\ \text{for } 2t < c.
\end{equation}
For $2t \geq c$, the interval $[\frac{c}{2}-t,\frac{c}{2}+t]$ contains both
possible values of $X$ and thus we have
\begin{equation}\label{eq:conc_b_2}
    Q_{X}(t) = 1,\ \text{for } 2t \geq c.
\end{equation}

\section{From infinite magic to type III factors}\label{app:araki-woods}

\begingroup
We prove Theorem~\ref{thm:aw_magic} for the sequence of states
\begin{equation}
    \ket{\psi^N} = \bigotimes_{i=1}^{N} \ket{\psi_i},
\end{equation}
where $\ket{\psi_i} \in \bbC^2 \otimes \bbC^2$ and for each $i$ we call the first copy of $\bbC^2$ the A-system. Again, we consider the algebra $M_A$ as the bicommutant in the GNS representation that the sequence of states $\ket{\psi^N}$ induces on the algebra generated by operators acting on a finite number of qubits in $A$. See e.g. \cite{ArakiWoods1968, Witten:2018zxz, Chemissany:2025vye} for further details. We use the stabilizer fidelity,
locally optimized stabilizer fidelity, and locally optimized min-relative
entropy of magic defined in Eqs.~\eqref{eq:stabilizer_fidelity},
\eqref{eq:locally_optimized_stabilizer_fidelity}, and
\eqref{eq:locally_optimized_min_relative_entropy}. The local optimization is
needed because the type of the factor depends only on the Schmidt spectra, whereas
stabilizer fidelity is not invariant under arbitrary local basis changes. The
local optimization removes this basis dependence. We first prove that the
sequence of states has bounded magic iff the resulting algebras are not type
III.
We then prove that the sequence of states has divergent magic iff it gives rise
to a type III algebra.
\subsection{Type I or II implies bounded magic}
    Let us Schmidt-decompose $\ket{\psi_i}$ as
    \begin{equation}
        \ket{\psi_i} = \sqrt{p_i} \ket{i_1 i_1} + \sqrt{1-p_i} \ket{i_2 i_2}.
    \end{equation}
    The reduced density matrix of the state $\ket{\psi^N}$ on the system $A$ then has the spectrum
    \begin{equation}\label{eq:finite_spectrum}
        \l^{(N)}_s = \prod_{i=1}^{N}p_i^{\s_i} (1-p_i)^{1-\s_i},
    \end{equation}
    where $s \in \{0,1 \}^N$ has $i$'th entry $\s_i$. In the following we denote by $a_j = \mathrm{max}(p_j,1-p_j)$ the largest eigenvalue of the $j$'th state and by $c_j = 1-a_j$ its smaller eigenvalue.
    Lemma 2.14 in Ref.~\cite{ArakiWoods1968} yields the following criteria for the type of the von Neumann algebra $M_A$
   \begin{itemize}
       \item[1)] $M_A$ is type I iff
           \begin{equation}
               \sum_j c_j < \infty. \label{eq:type-1}
           \end{equation}
       \item[2)] $M_A$ is type II$_1$ iff
           \begin{equation}
               \sum_j |\frac{1}{2}-p_j|^2 < \infty. \label{eq:type-2}
           \end{equation}
        \item[3)] $M_A$ is type III iff
    \begin{equation}
        \sum_{j=1}^{\infty} c_j \mathrm{min}\{|\frac{a_j}{c_j}-1|^2, C\} = \infty, \label{eq:type-3}
    \end{equation}
    where we set the summand to zero if $c_j =0$ and $C$ is an arbitrary constant,
that we set to $C = 1$ in the following. 
   \end{itemize}
   We now demonstrate that either of those conditions imply that the sequence
   of vectors has bounded magic.
We can choose local unitaries $U_A,U_B$ such that on the $i$'th tensor factor one has
\begin{equation}
    (U_A\otimes U_B) \ket{\psi_i} = \sqrt{a_i} \ket{00} + \sqrt{c_i} \ket{11}.
\end{equation}
Since the fidelity is a supremum over stabilizer states, we can lower bound it by
taking the overlap with a particular example. 
If the algebra is type II$_1$, we consider a tensor product of $N$ Bell states
\begin{equation}
    \ket{\p} = \frac{1}{\sqrt{2}} (\ket{00} + \ket{11}).\label{eq:bell-state}
\end{equation}
Taking the overlap of $N$ tensor products of these vectors then implies
\begin{equation}
    F_{\mathrm{STAB},\mathrm{LO}}^{(N)}(\psi^N) \geq \prod_{j=1}^{N}
    \frac{(\sqrt{a_j}+\sqrt{c_j})^2}{2} = \prod_{j=1}^{N}b_j,
\end{equation}
where we defined 
\begin{equation}\label{eq:def_bj}
    b_j = \frac{(\sqrt{a_j} + \sqrt{c_j})^2}{2} =
    \frac{1}{2}+\sqrt{a_j c_j}.
\end{equation}
We can now introduce
\begin{equation}
    \d_j = p_j - \frac{1}{2},\label{eq:def-dj}
\end{equation}
for which it follows 
\begin{equation}\label{eq:relation_bj_dj}
    1-b_j = \frac{2\d_j^2}{1+\sqrt{1-4\d_j^2}}.
\end{equation}
From the definition of $\d_j$ it follows that
\begin{equation}
    -\frac{1}{2} \leq \d_j \leq \frac{1}{2}.
\end{equation}
We can now conclude that
\begin{equation}\label{eq:bounding_bj_dj}
    \d_j^2 \leq 1-b_j \leq 2\d_j^2,
\end{equation}
where the lower bound follows from setting the denominator in
\eqref{eq:relation_bj_dj} to $2$ and the upper bound from setting it to $1$.
It is a standard fact that for $b_j \geq C > 0$ one has
\begin{equation}
    \prod_{j=1}^{\infty} b_j > 0 \Leftrightarrow \sum_{j=1}^\infty
    (1-b_j)
    < \infty.\label{eq:sum-prod-exchange}
\end{equation}
In our case we can set $C = \frac{1}{2}$ due to~\eqref{eq:def_bj}. From
\eqref{eq:bounding_bj_dj} we see that the sum is convergent iff 
\begin{equation}
    \sum_j \d_j^2 = \sum_j (\frac{1}{2}-p_j)^2 < \infty,
\end{equation}
which is equivalent to the type II criterion. In the type I case, instead of
lower bounding the fidelity by the overlap with the Bell states $\p$, we lower
bound it with the overlap with the product state $\ket{00}$, from which we
find 
\begin{equation}
    \inf_{N} F_{\mathrm{STAB},\mathrm{LO}}^{(N)}(\psi^N) \geq \prod_{j=1}^{\infty} a_j
\end{equation}
Repeating the same argument as before then leads to
\begin{equation}
    \prod_{j=1}^{\infty} a_j > 0 \Leftrightarrow \sum_j c_j < \infty,
\end{equation}
but this is the criterion for a type I algebra to emerge. 
If $M_A$ has type II$_\infty$, then properties~\eqref{eq:type-1}-\eqref{eq:type-3} fail. We can infer finite magic already from just the failure of Eq.~\eqref{eq:type-3}. In this criterion the summand $h_j = c_j \min\{|\frac{a_j}{c_j}-1|^2,1\}$ switches behavior at $c_j = \frac{1}{3}$. We will now repeat the analysis from the last two cases and take the overlap with stabilizer states depending on the value of $c_j$, i.e., we lower bound 
\begin{equation}
F_{\mathrm{STAB},\mathrm{LO}}^{(N)}(\psi^N) \geq \prod_{j = 1}^N |\langle \c_j |\psi_j \rangle|^2,
\end{equation}
where we pick $\ket{\c_j}$ as follows: If $c_j\leq\frac{1}{3}$, then the vector is closer to the type I case and we take the overlap with the product state $\ket{\c_j} = \ket{00}$. In this case the overlap is $|\langle \c_j | \psi_j\rangle |^2= a_j = 1-h_j$, where the last equality follows by direct calculation of $h_j$ for $c_j < \frac{1}{3}$. Thus we have that $1-|\langle \c | \psi_j\rangle |^2 = h_j$.
If $c_j > \frac{1}{3}$ we take the overlap with the Bell state $\ket{\c_j} = \ket{\phi}$ from Eq.~\eqref{eq:bell-state}. Then we have that $|\langle \c_j | \psi_j\rangle|^2 = b_j$ with $b_j$ defined in Eq. \eqref{eq:def_bj}. Eq. \eqref{eq:bounding_bj_dj} already demonstrates that $1-|\langle \c | \psi_j\rangle |^2  = 1-b_j \leq 2 \d_j^2$ with $\d_j$ defined in Eq.~\eqref{eq:def-dj}. By direct computation one can check that $h_j = 4 \frac{\d_j^2}{c_j}$. Since $c_j < 1$ we thus have $1-|\langle \c | \psi_j\rangle |^2 \leq h_j$. 
Therefore the failure of Eq.~\eqref{eq:type-3} implies 
\begin{equation}
    \sum_j 1-|\langle \c | \psi_j\rangle |^2 \leq \sum_j h_j < \infty.
\end{equation}
In both cases we have chosen the vectors such that $|\langle \c | \psi_j\rangle |^2 \geq \frac{1}{2}$. Thus we can apply Eq.~\eqref{eq:sum-prod-exchange} and conclude
\begin{equation}
     \inf_{N} F_{\mathrm{STAB},\mathrm{LO}}^{(N)}(\psi^N) \geq \prod_{j = 1}^\infty |\langle \c_j |\psi_j \rangle|^2 > 0,
\end{equation}
which again proves finite magic in the case of type II$_\infty$. Note that this last proof only used the failure of the type III criterion and thus already implies the boundedness of magic in the type I and II$_1$ case.
We have shown that both type I and type II imply a nonzero limit for the locally optimized stabilizer fidelity and in turn a finite locally optimized min-relative entropy of magic, i.e., 
\begin{equation}
    \inf_{N\geq 1} F_{\mathrm{STAB},\mathrm{LO}}^{(N)}(\psi^N)>0 \ \text{if }M_A\text{ is type I or II.}
\end{equation}

\subsection{Type III implies divergent magic}
We will now prove that in the case of type III, the magic measure $\cD^{(N)}_{\mathrm{min,LO}}(\psi^N)$ diverges. To that end we will upper bound the stabilizer fidelity and show that the upper bound asymptotically vanishes. 
It is easy to see that for $r \geq 1$ 
\begin{equation}\label{eq:ineq_log_min}
    \mathrm{min}\{(\log r)^2,1 \} \geq \frac{1}{4} \mathrm{min}\{(r-1)^2, 1\}.
\end{equation}

    We see that Eq.~\eqref{eq:ineq_log_min} implies that in this case one also
    has that 
    \begin{lemma}[Type III and divergent log-spectrum fluctuations] If the sequence of states $\psi^N$ is such that $M_A$ has type III, i.e.,~\eqref{eq:type-3} holds, then
        \begin{equation}\label{eq:divergent_fluctuations}
           \lim_{N \rightarrow \infty} \sum_{j = 1}^{N} c_j \mathrm{min}\{(\log
           \frac{a_j}{c_j})^2,1 \} = \infty.
        \end{equation}
        Terms with $c_j = 0$ are set to $0$ in this sum.
    \end{lemma}
    This will be the criterion we will assume to judge that the limiting algebra has type III.

\subsubsection{Bounding stabilizer fidelity}
    The spectrum of the reduced state at $n$ layers is given by the numbers 
    \begin{equation}\label{eq:lambda_s_l}
        \l_s^{(n)} = \prod_{j=1}^{n} a_j^{1-s_j} c_j^{s_j},
    \end{equation}
    where $s \in \{0,1\}^n$ is a bit string and we set $0^0=1$. Each bit string gets mapped to
    a definite eigenvalue of the bipartite state. To demonstrate
    that type III structure forces the fidelity to vanish, we will construct an
    upper bound and prove that this upper bound asymptotically vanishes as $n
    \rightarrow \infty$. For this we need
    
    \begin{lemma}[Stabilizer fidelity bound from the Schmidt spectrum]
        Let $\varnothing\neq S \subset \{0,1\}^{n}$ and 
        \begin{equation}
            A_n(S) = \frac{1}{\sqrt{|S|}} \sum_{s\in S} \sqrt{\l^{(n)}_{s}},
        \end{equation}
        then
        \begin{equation}\label{eq:F_leq_A}
            F_{\mathrm{STAB},\mathrm{LO}}^{(n)}(\psi^n) \leq \big(\sup_{\varnothing\neq S\subset \{0,1 \}^n}
            A_n(S)\big)^2
        \end{equation}
    \end{lemma}
    
    \begin{proof}
        Any stabilizer state $\chi$ has a flat bipartite spectrum \cite{Fattal:2004frh}
        \begin{equation}
            \m_i = \begin{cases}
                1/r,\qquad 1 \leq i \leq r,\\
                0,\qquad i > r,
            \end{cases}
            \qquad r = 2^{k},
        \end{equation}
        for some $0 \leq k \leq n$.
        Using the Schmidt decomposition of $\ket{\psi^n}$, one can show that \cite{Viscardi:2026oos}
        \begin{equation}
            \sup_{U_A,U_B} |\bra{\c}U_A \otimes U_B \ket{\psi^n}| =
            \frac{1}{\sqrt{r}} \sum_{i=1}^{r} \sqrt{\l_i},
        \end{equation}
        where we have ordered the spectrum of $\psi^n$ in descending order, i.e.,
        $\l_1$ is the largest eigenvalue, $\l_2$ the second largest etc.
        We therefore find that for any stabilizer state
        \begin{equation}\label{eq:upper_bound_fidelity}
            \sup_{U_A,U_B} |\bra{\c} U_A \otimes U_B \ket{\psi^n}| \leq
                    \sup_{\varnothing\neq S \subset \{0,1 \}^n} A_n(S).
        \end{equation}
        Since the fidelity is the square of the supremum of the above over all stabilizer states and the
        above inequality is true for every stabilizer state, Eq.
       ~\eqref{eq:F_leq_A} follows.
    \end{proof}

    \subsubsection{Type III and anti-concentration}
    The above discussion demonstrates that it is useful to think of the
    eigenvalue distribution as providing a probability distribution on random
    bitstrings $s \in \{0,1 \}^n$.
    Here we prove that in the case of a type III algebra, this probability
    distribution will not concentrate at some particular value. This is 
    exactly the statement of \emph{antiflatness} \cite{Tirrito:2023fnw} and here we will
    demonstrate that sequences of vectors leading to type III factors exhibit a strong form
    of antiflatness.
    That is, we consider a random string $S_n$ with the probability
    of $S_n = s$ being given by 
    \begin{equation}
        P(S_n = s) = \l_s^{(n)}.
    \end{equation}
    We can then consider the function on the positive-probability strings
    \begin{equation}
        X_n(s)= - \log \l_s^{(n)} = -\sum_{j = 1}^{n} \log a_j + \sum_{j=1} s_j
        \log \frac{a_j}{c_j}.\label{eq:def-log-spectrum}
    \end{equation}
    This can be written as
    \begin{equation}
        X_n = C_n + \sum_{j=1}^{n} d_j B_j,
    \end{equation}
    with $C_n=-\sum_{j=1}^n\log a_j$ and $d_j = \log(a_j/c_j)$ for $c_j>0$, while $d_j=0$ if $c_j=0$; $B_j$ are independent Bernoulli random variables with
    \begin{equation}
        \mathbb{P}(B_j = 1) = c_j, \qquad \mathbb{P}(B_j = 0) = a_j.
    \end{equation}
    These identities for $X_n$ hold almost surely; bit strings with $\lambda_s^{(n)}=0$ have zero probability. The statement of antiflatness then becomes the question whether the random
    variable concentrates on a particular value. In contrast we have in the type III case 
    \begin{lemma}[Anti-concentration of the $X_n$] \label{lem:anti_concentration}
        Let $X_n$ be the sequence of random variables defined in Eq.~\eqref{eq:def-log-spectrum} with the spectrum arising from a sequence of states $\psi^N$ for which $M_A$ is a type III factor. Then for every fixed $L > 0 $ the concentration function asymptotically vanishes, i.e., 
        \begin{equation}
           \lim_{n\to \infty} Q_{X_n}(L) = 0.
        \end{equation}
        
    \end{lemma}
    \begin{proof}
        We will restrict our attention to $\tilde{X}_n := \sum_{j \in I_n} d_j
        B_j$, where $I_n$ is the set of indices for which $d_j \neq 0$. This is
        valid as the concentration functions of random variables shifted by constants
        agree, since the concentration function takes the supremum over all values.
        For every $j \in I_n$, set
        \begin{equation}
            \ell_j := \frac{1}{4} \mathrm{min}\{d_j,L,1\}.
        \end{equation}
        This will become the evaluation point in the Kolmogorov-Rogozin
        inequality. It is valid, since $\ell_j < L$, so we have 
        \begin{equation}
            Q_{\tilde{X}_n}(L) \leq \frac{C L}{\sqrt{\sum_{i\in I_n}
            \ell^2_i (1-Q_{d_i B_i}(\ell_i))}}.
        \end{equation}
        Since $\ell_j < d_j/4$, Eq.~\eqref{eq:conc_b_1} implies that 
        $Q_{d_j B_j}(\ell_j) = a_j$ so that 
        \begin{equation}\label{eq:Q_upper_bound_1}
            Q_{\tilde{X}_n}(L) \leq \frac{C L}{\sqrt{\sum_{i\in I_n}
            \ell^2_i c_i}}.
        \end{equation}
        We furthermore have that 
        \begin{equation}\label{eq:ell_j_ineq}
            \ell_j = \frac{1}{4} \mathrm{min}\{d_j,L,1\} \geq \frac{1}{4}
            \mathrm{min}\{L,1\} \mathrm{min}\{d_j,1\}.
        \end{equation}
        This holds because if $\ell_j = \frac{1}{4}$, then both $d_j \geq 1$ and $L \geq 1$
        and the inequality is $\frac{1}{4} \geq \frac{1}{4}$, which is true.
        If $\ell_j < \frac{1}{4}$, then at least one of $d_j$ or $L$ is smaller than $1$. If only one is
        smaller than $1$, then one of the two minima on the r.h.s. equal one
        and the other equals $4 \ell_j$, so again one has $\ell_j \geq \ell_j$.
        If both are smaller than one, then one has
        \begin{equation}
            \ell_j \geq \ell_j \mathrm{max}(d_j,L),
        \end{equation}
        which is true, since we now assumed the arguments in the maximum to be
        less than one, so we multiply $\ell_j$ with a number $\leq 1$.
        Squaring Eq.~\eqref{eq:ell_j_ineq} and inserting it into Eq.~\eqref{eq:Q_upper_bound_1}, we find 
        \begin{equation}
            Q_{\tilde{X}_n}(L) \leq \frac{C L}{\sqrt{\sum_{j\in I_n} \frac{1}{16}
            \mathrm{min}\{L^2,1\} \mathrm{min}\{d_j^2,1\}c_j}} = \frac{4 C L
        }{\sqrt{\mathrm{min}\{L^2,1\} V_n}},
        \end{equation}
        where $V_n$ is the sum in Eq.~\eqref{eq:divergent_fluctuations}. The
        same equation also shows that $V_n \rightarrow \infty$ and therefore
        \begin{equation}
           \lim_{n \rightarrow \infty} Q_{\tilde{X}_n}(L) = 0.
        \end{equation}
    \end{proof}
    
    \subsubsection{Type III implies asymptotically vanishing stabilizer fidelity}
        
    We now want to show that in the type III case, the upper bound in Eq.~\eqref{eq:upper_bound_fidelity} vanishes
    asymptotically.
    
    \begin{lemma}[Asymptotically vanishing optimized stabilizer-fidelity]
        \begin{equation}
          \lim_{n \rightarrow \infty}  \sup_{\varnothing\neq S \subset \{0,1 \}^n} A_n(S) = 0.
        \end{equation}
    \end{lemma}

    \begin{proof}
        We will first construct a uniform bound for $A_n(S)$ for any $S$ and
        prove that it vanishes asymptotically. We will achieve this by
        fixing an $R>0$ and splitting any $S \subset \{0,1 \}^{n}$ into three sets that quantify
        the deviation of $X_n$ from the flat case $X_n = \log |S|$ by at least
        a distance $R$. The contribution from each set $\S$ to $A_n(S)$ is 
        \begin{equation}
            C_\S = \frac{1}{\sqrt{|S|}} \sum_{s \in \S} \sqrt{\l^{(n)}_s},
        \end{equation}
        which we can estimate as follows.
        \begin{itemize}
            \item[1)] The Cauchy-Schwarz inequality and $|\S| \leq |S|$ implies 
        \begin{equation}\label{eq:crit_1}
            C_\S \leq \sqrt{\frac{|\S|}{|S|}}  \sqrt{\sum_{s \in \S}
            \l^{(n)}_s} =
            \sqrt{\frac{|\S|}{|S|}} \sqrt{\mathbb{P}(s \in \S)} \leq
            \mathrm{min}\{\sqrt{\frac{|\S|}{|S|}},\sqrt{\mathbb{P}(s \in \S)}
            \},
        \end{equation}
        where the last inequality holds, because both factors are $\leq 1$.
            \item[2)] Defining $\l_\S = \sup_{\l_s: s \in \S} \l_s$ when $\S\neq\varnothing$ (and taking $\lambda_\varnothing=C_\varnothing=0$) one has
                \begin{equation}\label{eq:crit_2}
                    C_\S \leq \frac{|\S|}{\sqrt{|S|}} \sqrt{\l_\S} \leq
                    \sqrt{|S| \l_\S},
                \end{equation}
                where the second inequality again used $|\S| \leq |S|$.
        \end{itemize}
        Now for any fixed $S \subset \{0,1 \}^{n}$, the flat case has
        \(X_n = \log |S|\). We now split the set $S$ into the subsets
        \begin{equation}
          S_-:=\{s\in S:X_n(s)<\log|S|-R\},
        \end{equation}
        \begin{equation}
          S_0:=\{s\in S:\log|S|-R\le X_n(s)\le \log|S|+R\},
        \end{equation}
        \begin{equation}
          S_+:=\{s\in S:X_n(s)>\log|S|+R\}.
        \end{equation}

        On \(S_-\), we have
        \begin{equation}
          \lambda_s^{(n)}=e^{-X_n(s)}>e^R|S|^{-1}.
        \end{equation}
        Since \(\lambda^{(n)}\) is a probability distribution, one has
        \begin{equation}
          1
          \ge
          \sum_{s\in S_-}\lambda_s^{(n)}
          >
          |S_-|\,e^R|S|^{-1}.
        \end{equation}
        Thus
        \begin{equation}
          |S_-|\le |S|e^{-R}.
        \end{equation}
        By Eq.~\eqref{eq:crit_1} one thus has
        \begin{align}
          C_{S_-}\le
          e^{-R/2}.
        \end{align}

        On \(S_+\), we have
        \begin{equation}
          \lambda_s^{(n)}<e^{-R}|S|^{-1}.
        \end{equation}
        Therefore Eq.~\eqref{eq:crit_2} gives
        \begin{equation}
          C_{S_+} \le  e^{-R/2}.
        \end{equation}

        On \(S_0\) we use~\eqref{eq:crit_1}, which gives
        \begin{equation}
            C_{S_0} \leq \sqrt{\mathbb P(s \in S_0)}.
        \end{equation}
        By construction, $S_0$ is the set for which $X_n$ lies in the
        $R$-sized window $[\log|S|-R,\log|S|+R]$ around the flat case. Therefore $\mathbb{P}(s \in S_0)$ is by definition bound by its
        concentration function in Definition~\ref{def:concentration_function}.
        Hence
          \begin{equation}
              C_{S_0} \leq \sqrt{Q_{X_n}(R)}
          \end{equation}
        Combining the three estimates we then have
        \begin{equation}
          A_n(S)
          \le
          2e^{-R/2}+\sqrt{Q_{X_n}(R)}.
        \end{equation}
        The bound is uniform in the set $S$ and thus bounds the supremum of
        $A_n(S)$ for a fixed $n$. For fixed $R$, the anti-concentration
        lemma \ref{lem:anti_concentration} gives $Q_{X_n}(R)\to 0$ so that          
        \begin{equation}
          \lim_{n\to\infty}
          \sup_{\varnothing\neq S\subseteq\{0,1\}^n} A_n(S)
          \leq 2e^{-R/2}.
        \end{equation}
        Since this holds for every $R>0$, we have
        \begin{equation}
          \lim_{n\to\infty}
          \sup_{\varnothing\neq S\subseteq\{0,1\}^n} A_n(S)
          =
          0.
        \end{equation}
    \end{proof}
\endgroup

\stoptoc
\twocolumngrid

\bibliography{references,BigReferences70}

\end{document}